\documentclass[nofootinbib,preprint,aps,prd]{revtex4-1}
\usepackage{graphicx}
\usepackage{amssymb}
\usepackage{slashed}
\newcommand{\be}{\begin{eqnarray}}
\newcommand{\ee}{\end{eqnarray}}
\usepackage[colorlinks]{hyperref}
\hypersetup{
	colorlinks=true,
	citecolor=blue,
	filecolor=magenta,      
	urlcolor=cyan,
}
\newcommand{\nn}{\nonumber}

\usepackage{amsmath,amssymb}
\usepackage{mathtools}
\usepackage{float}
\usepackage{color}
\usepackage{leftidx}
\usepackage{booktabs}
\usepackage{subcaption}
\usepackage{amsfonts}
\usepackage{multirow}
\usepackage[utf8]{inputenc}
\newcommand{\mathsym}[1]{{}}
\newcommand{\unicode}[1]{{}}

\begin{document}
	
	\title{Reheating constraints to modulus mass for single field inflationary 
		models}
	
	\author{Rajesh Goswami}
	\email{rajesh@phy.iitb.ac.in}
	\author{ Urjit A. Yajnik}
	\email{yajnik@phy.iitb.ac.in}
	\address{ Department of Physics,
		Indian Institute of Technology Bombay, Mumbai-400076,
		India}

\begin{abstract}
We consider string and supergravity motivated scenarios in which moduli fields 
dominate the energy density of the Universe in a post-inflationary epoch. 
For the case of  a single light modulus it has been shown that considering the evolution of a specific scale from the time 
of its Hubble crossing during inflation to the present time, a  relation can be obtained among the lightest modulus mass, the 
reheating parameters ($T_{\text{reh}}$, $\bar{w}_{\text{reh}}$ and $N_{\text{reh}}$) and the  inflationary observables. 
By paying closer attention to the role of the $\bar{w}_{\text{reh}}$, we obtain more stringent 
constraints on the value of the modulus mass and the reheating parameters using the CMB data. 
Next, the analysis is extended  to include features in the inflaton potential as a source of CMB low multipole anomalies, which 
further constrains the mass of the modulus to be substantially higher than 
without such a constraint. By both considerations and for several inflation models considered, we find a
constraint on the mass of the lightest modulus particle, $m_{\chi}$, generically $\gtrsim10^{15}$GeV, with possible 
low values $\sim10^{12}$GeV. While a simplification of the reheating phase is assumed, the bounds are reliably suggestive,
and the study may be taken  as a demonstration that substantial  knowledge about reheating phase buried deep in 
the early epochs of the Universe is accessible through the use of CMB observables today.
\end{abstract}
\maketitle
\raggedbottom

\section{Introduction}\label{sec1}
The data from the recently concluded Cosmic Microwave Background (CMB) experiments 
\cite{Aghanim:2018eyx, Akrami:2018odb, Ade:2015xua, Ade:2015lrj, 
Planck:2013jfk, Ade:2013zuv, Dunkley:2008ie,Komatsu:2008hk, 
0067-0049-192-2-18,0067-0049-208-2-19, Peiris:2003ff,0067-0049-192-2-16,
0004-637X-705-1-978, Reichardt:2008ay} are in 
perfect agreement with the scale invariant spectrum  
\cite{Guth:1982ec, Starobinsky:1982ee, Mukhanov:1990me, Bardeen:1983qw,
Riotto:2002yw}  as predicted by the theory of inflation 
\cite{Starobinsky:1980te, Guth:1980zm, Linde:1981mu, Hawking:1982cz, 
Linde:1983gd, Linde:2007fr, Martin:2013tda}, and this puts forth slow-roll inflation as 
the leading candidate for the early Universe cosmology. The slow-roll conditions 
of inflation are ultraviolet sensitive, and one should embed inflationary 
models in quantum theory of gravity. String theoretical models of inflation take 
care of these ultraviolet issues. However, in string or supergravity models 
\cite{Burgess:2013sla, Silverstein:2013wua, Baumann:2014nda}, there are moduli 
fields which play a central role. These are generic scalar fields
which are massless in the basic construction, but acquire masses much lighter than
the string scale through subleading corrections. 
At the end of the inflationary phase, the inflaton field oscillates 
and brings the Universe to thermal equilibrium; this phase is generically 
referred to as reheating \cite{Turner:1983he, Traschen:1990sw, Albrecht:1982mp, 
Kofman:1994rk, Kofman:1997yn, Drewes:2013iaa, Allahverdi:2010xz}. 
The presence of moduli whose masses are lighter than the value of the Hubble parameter after
the Universe returns to a thermal equilibrium post inflation are important to subsequent cosmology,
while the heavier ones can be considered to have been inflated away. 
The light  moduli with almost flat potentials are displaced from their minima during 
inflation, but subsequently begin to oscillate and manifest as light particles.
In this paper we consider the case of a single modulus field as an example, and 
consider it in conjunction with a duly parametrised reheating scenario.

According to the standard cosmology, the early Universe passed through the 
following epochs: inflation, reheating, radiation domination and matter 
domination. Now, if the Universe becomes modulus dominated after the radiation 
dominated epoch and reheats the Universe for the second time, then the Universe 
has gone through the epochs : inflation, reheating (inflaton decay), radiation 
dominated, modulus dominated, reheating (modulus decay) and matter
dominated eras. In this paper, we relate the reheating and inflationary 
parameters to the lightest modulus mass, ($m_{\chi}$), by considering the 
second reheating phase (modulus decay) is instantaneous, and obtain tight 
constraints on the modulus mass as well as on the reheating parameters. 
Early treatments of this question have come to the conclusion that $m_{\chi}>3\times10^{4}$GeV 
\cite{Coughlan:1983ci, Banks:1993en, deCarlos:1993wie} while a recent study \cite{Dutta:2014tya}  
which follows the same methodology  as adopted here obtains a more stringent bound $m_{\chi}>10^9$GeV. 
However, extending our previous work  \cite{Goswami:2018vtp} we treat the effective equation of 
state parameter during reheating, $\bar{w}_{\text{reh}}$ in more detail. Further, using the Planck 2018 range
of values for $n_s$ we are able to obtain much 
more stringent bound on the $m_{\chi}$, generically $\gtrsim10^{15}$GeV, but with a weaker lower bound $\sim10^{12}$GeV. 

As another input, we consider the fact that at lower multipoles, specifically around 
$\ell={\bf{22}}$ and $\bf{40}$, the Planck data points lie outside the cosmic variance associated with the power law 
primordial spectrum. If not of a completely accidental origin, this raises the possibility of non-trivial inflationary dynamics,
in turn providing important phenomenological inputs about the inflationary  model. Model independent approaches
to reconstruct the primordial power spectrum from the CMB anisotropies have been reported
in \cite{Hannestad:2000pm,Bridle:2003sa,Mukherjee:2003ag,Hannestad:2003zs,Shafieloo:2003gf,Shafieloo:2006hs,
Shafieloo:2007tk,Nagata:2008zj,1475-7516-2009-07-011}. 
The consideration of a burst of oscillations in the primordial power spectrum leads to a good fit to the 
CMB angular power spectrum, including the anomalies \cite{Hazra:2010ve, Hazra:2016fkm,Hazra:2017joc}. 
In order to generate these oscillations in 
the primordial power spectrum, one has to consider a short 
period of deviation from slow-roll inflation \cite{Starobinsky:1992ts,Dvorkin:2009ne}. A possible approach
to such a deviation is to introduce a step in the inflaton 
potential \cite{Adams:2001vc, Covi:2006ci, Hamann:2007pa, Mortonson:2009qv, Joy:2007na, Jain:2008dw, 
Jain:2009pm}. A step with suitable height and width at a particular location of the inflationary potential 
has resulted in a better fit to the CMB data near the multipole $\ell={\bf 22}$. In  \cite{Garg:2018trf} 
a possible origin for an unusual phase at the onset of inflation that can produce such a deviation has been 
considered in the context of $SO(10)$ grand unification. 

It can be shown that the generic relation between late time observables and reheating phase 
in a single field inflation can be strengthened by demanding successful explanation of the low 
multipole anomalies. The link is the specific position, $\phi_k/M_{Pl}$, of the inflaton in the course of its slow 
roll, at which it encounters the step in the potential. The location of such a step in the inflaton potential 
was obtained in Ref. \cite{Hazra:2010ve}. Then it can be shown \cite{Goswami:2018vtp} that such 
a step makes the constraints on the reheating parameters more stringent  for different inflationary models. 
In this paper we apply this method in conjunction with a single late time light modulus. This approach
also gives much higher lower bound on $m_{\chi}$, $\gtrsim10^{12}$GeV.
 
This article is organized as follows: Sec \ref{sec2intro} begins with emphasising the crucial role of the 
reheating phase and the relatively constrained nature of the parameters governing it.
Its subsection \ref{sec2} deals with the slow-roll 
inflation and late time modulus dominated cosmology. In this and the following subsection we derive 
the relation of $m_{\chi}$ with $T_{\text{reh}}$, $N_{\text{reh}}$, 
$w_{\text{reh}}$ and the inflationary parameters ($V_{\text{end}}$ and $\Delta 
N_{k}$). The expression for $m_{\chi}$ is derived as a function of the scalar 
spectral index $n_{s}$ for different single field inflationary models in Sec 
\ref{sec3}. Sec. \ref{Sec4} studies the additional constraint on the modulus mass 
$m_{\chi}$ due to the addition of a step in the potential along the lines of \cite{Goswami:2018vtp}. 
Finally, Sec \ref{Sec6} contains the conclusions.

We work with $ \hbar=c=1$ units and the following values are used. 
$M_{\text{Pl}}=\sqrt\frac{1}{8\pi G} 
=2.435\times10^{18}\text{GeV}$ is the
reduced Planck mass and  the redshift of last scattering surface is 
$z_{\text{ls}}=1100$. The $ z_{\text{eq(MR)}}=3402$ 
is the redshift of matter radiation 
equality and the present value of the Hubble parameter $ H_{0}=100 h$ km 
$\text{s}^{-1} \text{Mpc}^{-1} $ with
$h=0.6736$ \cite{Aghanim:2018eyx}.

\section{Connecting inflation parameters to observable scales}
\label{sec2intro}
Reheating phase of inflationary models may appear to not possess any observable, however the almost scale
invariant nature of matter perturbations in the late Universe would require more complicated explanations, were the
transitional reheating phase also not essentially of Friedmann type. 
Thus one can parametrize the reheating phase as, the temperature of reheating 
($T_{\text{reh}}$), the duration of reheating ($N_{\text{reh}}$), and a 
somewhat coarse but useful averaged equation of state parameter during reheating ($\bar{w}_{\text{reh}}$).
As we shall see, due to the limited range of variation available to this parameter $w_{\text{reh}}$ on
physical grounds, it ends up providing strong constraints on the global inflationary scenario.
Although the reheating parameters seem to be hopelessly far 
away from being observationally determined, the natural range for 
$\bar{w}_{\text{reh}}$ being limited as $-\frac{1}{3}\leq\bar{w}_{\text{reh}}\leq 1$
provides substantial constraints on the rest of the reheating scenario.

After reheating, the Universe becomes radiation dominated. The expansion of the 
Universe redshifts the energy density associated with it, and the Hubble 
parameter ($H$) value decreases. When $H$ becomes comparable with the mass of 
the light moduli, the moduli fields start oscillating around 
the minimum of their respective potentials \cite{Coughlan:1983ci,Banks:1993en, 
deCarlos:1993wie, Kane:2015jia}. The 
energy associated with the moduli dilutes 
like matter which is at a rate slower than the radiation. Hence, very quickly 
the energy density of the Universe becomes modulus dominated. Ultimately, the 
moduli decay, and as a consequence, the Universe should reheat for the second 
time. In ref.s \cite{Dutta:2014tya, Das:2015uwa, Cicoli:2016olq,
Bhattacharya:2017ysa, Bhattacharya:2017pws, Maharana:2017fui} 
string inflation was studied with the presence of moduli implying constraints 
either on inflation parameters or on the moduli potential and on the lightest 
modulus mass. In this paper we consider the case of single light modulus field 
of mass $m_{\chi}$. Due to the prompt decay of the moduli, we can consider the 
second reheating phase to be short enough that it may be treated as 
instantaneous.
We analyse the consequences of this sequence of events in detail, keeping in 
mind the natural restrictions on the possible values of the reheat parameters. 
The related essential formalism concerning inflation which is now fairly standard is 
recapitulated in the appendix \ref{sec2}.  There it is shown that the  temperature after 
the modulus decay can be  written in terms of the modulus mass as 
\be\label{ch9reh}
T_{\text{decay}}\sim m_{\chi}^{3/2}M_{\text{Pl}}^{-1/2}
\ee
The lower bound of the reheat temperature is around a few MeV (the BBN 
temperature). Hence, using Eq. \eqref{ch9reh} one obtains the bound on the 
modulus mass (known as cosmological moduli problem bound) as 
\cite{Coughlan:1983ci,Banks:1993en,deCarlos:1993wie}
\be
m_{\chi}\ge30 \text{TeV}
\ee
However in our analysis we will find that constraint 
on the $m_{\chi}$ is substantially different from this because we have traced the
dependence on $w_{\text{reh}}$ in detail. Later we shall recapitulate the source of our difference with the previous 
work.

\begin{figure}[bht]
 \centering
  \includegraphics[width=1\linewidth]{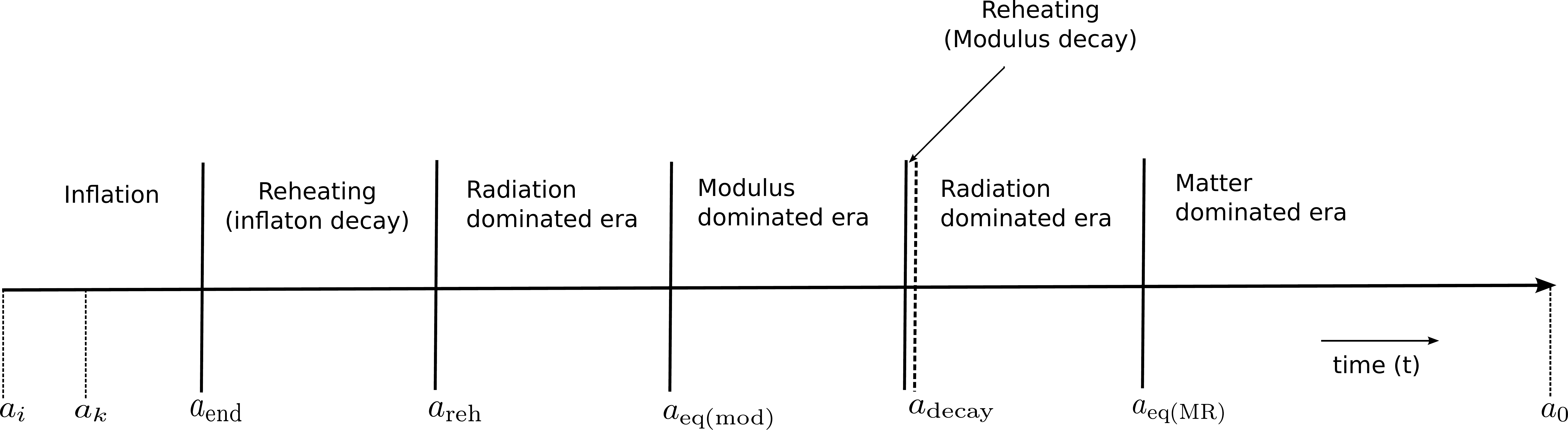}
\caption[Era]{A non-standard evolution of
our Universe, which consists of the following epochs -- inflation, reheating 
(inflaton decay), radiation domination, modulus domination, reheating due to 
modulus decay (we consider this epoch instantaneous which is shown in the narrow 
region), radiation domination and matter domination (we ignore the very 
recent cosmological constant ($\Lambda$) domination). Here $a_{i}$, $a_{k},~ 
a_{\text{end}}, ~ a_{\text{reh}}, ~ a_{\text{eq(mod)}}, ~ 
a_{\text{decay}}, ~ a_{\text{eq(MR)}} ~\text{and}~ a_{0}$ represents the value 
of the scale factor at the beginning of inflation, Hubble 
crossing of a specific scale $k$, end of inflation, end of 
reheating (decay of inflaton), modulus radiation equality, end of modulus 
decay, matter radiation 
equality and at the present time respectively.}
 \label{figEra}
\end{figure}

We now adapt our method of \cite{Goswami:2018vtp} to show that  the mass of the lightest modulus
field can be related to cosmological observational parameters. This can be done by considering the evolution of a 
cosmological scale from the time of Hubble crossing during inflation to present 
time. While the discussion also parallels our earlier work, the specific expressions differ and for completeness we
include all the reasoning.
Fig. \ref{figEra} gives the various epochs we consider for ready reference.

A physical scale today $\frac{k}{a_0}$ can be related to it's value at the time 
of Hubble crossing during inflation, $\frac{k}{a_k}$, as
\be\label{Hk1}
\frac{k}{a_{k}}=H_{k}&=&\frac{k}{a_{0}}\frac{a_{0}}{a_{\text{eq(MR)}}}
\frac{a_{\text{eq(MR)}}}{a_{\text{decay}}}\frac{a_{\text{decay}}}{a_{
\text {eq(mod)}}}
\frac{a_{\text{eq(mod)}}}{a_{\text{reh}}}\frac{a_{\text{reh}}}{a_{\text{end}}}
\frac { a_ { \text{end} } } { a_ { k } },
\ee
where $a_{k},~ a_{\text{end}}, ~ a_{\text{reh}}, ~ a_{\text{eq(mod)}}, ~ 
a_{\text{decay}}, ~ a_{\text{eq(MR)}} ~\text{and}~ a_{0}$ represents the value 
of the scale factor at the time of Hubble crossing, end of inflation, end of 
reheating (decay of inflaton), modulus radiation equality, end of modulus 
decay, matter radiation 
equality and at the present time respectively. Throughout this paper, the 
subscripts ``reh" and ``decay" represent the end of reheating due to inflaton 
decay and modulus decay respectively. The Eq. \eqref{Hk1} can be written as
\be\label{Hk2}
H_{k}=\frac{k}{a_{0}}\left(1+Z_{\text{eq(MR)}}\right)\left(\frac{
\rho_{\text{decay}} } { \rho _{\text{eq(MR)}}} 
\right)^{1/4}e^{N_{\text{mod}}}e^{N_{\text{rad}}}e^{N_{\text{reh}}}e^{\Delta 
N_{k}}.
\ee
Here $Z_{\text{eq(MR)}}$ is the redshift of matter radiation equality, and 
$\Delta N_{k}$ indicates the number of e-folds remaining after the scale $k$ 
has crossed the Hubble radius during inflation. The quantity 
$N_{\text{mod}}$ and $N_{\text{rad}}$ represent the number 
of e-foldings in modulus dominated era and radiation dominated era (after 
inflaton decay). The energy density at the end of modulus 
decay and at the time of matter radiation equality are represented by 
$\rho_{\text{decay}}$ and $\rho _{\text{eq(MR)}}$ respectively. 
$N_{\text{reh}}$ 
is the number of e-folds during the period of reheating, in which an
epoch of preheating \cite{Boyanovsky:1996ks, Kofman:1997yn, Kofman:1997pt, 
Felder:1998vq,
Giudice:2001ep, Desroche:2005yt} is followed by
the thermalization process. Subsequent evolution of the Universe
governed by an energy density 
\be
\rho_{\text{reh}}=\frac{\pi^{2}}{30}g_{\text{reh}}T^{4}_{\text{reh}},
\ee
where $g_{\text{reh}}$ is the effective number of relativistic 
species, and $T_{\text{reh}}$ is the temperature at the end of 
inflaton decay.
The Eq. \eqref{Hk2} can be re written as 
\be
H_{k}=\frac{k}{a_{0}}\left(1+Z_{\text{eq(MR)}}\right)\left(\frac{
\rho_{\text{decay}} } { \rho _{\text{eq(MR)}}} 
\right)^{1/4} e^{N_\text{moddom}} 
\left(\frac{\rho_{\text{reh}}}{\rho_{\text{eq(mod)}}}\right) 
e^{N_{\text{reh}}}e^{\Delta N_{k}},
\ee
where $\rho_{\text{reh}}$ and $\rho_{\text{eq(mod)}}$ are energy density at the 
end of inflaton decay and at the time of modulus radiation equality 
respectively.
We can further parametrize the reheating phase (decay of inflaton) by 
considering that during that time the Universe was dominated by a fluid 
\cite{Martin:2003bt, Martin:2013tda}  of pressure P and energy density $\rho$, 
with an equation of state $w_{\text{reh}}=\frac{P}{\rho}$. Imposing the 
continuity 
equation, we have
\be
\dot{\rho}+3H(\rho+P)=0,\\
\dot{\rho}+3H\rho(1+w_{\text{reh}})=0.
\ee
In view of this equation, we have
\begin{eqnarray}
\label{e12}
\rho_{\text{reh}}&=&\rho_{\text{end}} e^{- 
3N_{\text{reh}}(1+\bar{w}_{\text{reh}})}; \\
\textrm{where} \qquad \bar{w}_{\text{reh}}=<w>&=&\frac{1}{ 
N_{\text{reh}}}\int_{N_{e}}^{N}{w_{\text{reh}}(N)}dN
\end{eqnarray}
Here $\bar{w}_{\text{reh}}$ is the 
average equation of state parameter during reheating \cite{Martin:2014nya}. 
From Eq. \eqref{e12}  we have 
\be\label{Hkenergy}
e^{N_{\text{reh}}}=\left(\frac{\rho_{\text{reh}}}{\rho_{\text{end}}}\right)^{
-\frac{1}{3(1+\bar{w}_{\text{reh}})}},
\ee
and the Eq. \eqref{Hk2} can now be rewritten as
\be
H_{k}=\frac{k}{a_{0}}\frac{a_{\text{decay}}}{a_{\text{eq(mod)}}}\rho_{\text{
decay } }
^ { 1/4 } \rho_ { \text{eq
(MR)}}^{-1/4}e^{N\text{mod}}\rho_{\text{reh}}^{\frac{3\bar{w}_{\text{reh}}-1}
{
12(1+\bar{w}_{\text{reh}})}}\rho_{\text{eq(mod)}}^{-1/4}\rho_{\text{end}}^{\frac
{1}{3(1+\bar{w}_{\text{reh}})}}e^{\Delta N_{k}}.
\ee
During the modulus dominated era, the energy density of the Universe scales 
as $\rho\sim a^{-3}$, and we can write 
\be\label{Nmod1}
e^{-N\text{mod}}=\left(\frac{\rho_{\text{decay}}}{\rho_{\text{eq(mod)}}}
\right)^{1/3+\alpha}.
\ee
During the epoch that the moduli dominate the energy density, $\alpha\approx0$, 
which is what we assume in the following.
Taking natural logarithm on both sides of Eq. \eqref{Nmod1}, one obtains
\be\label{ch9eq3}
-N_{\text{mod}}=\frac{1}{3}\ln\left(\frac{\rho_{\text{decay}}}{\rho_{\text{eq
(mod)} } } \right).
\ee
From Eq. \eqref{ch9eq3} we can write the energy density at the time of modulus 
radiation equality as 
\be\label{Nmod2}
\ln \rho_{\text{eq(mod)}}=\ln\rho_{\text{decay}}+3N_{\text{mod}}.
\ee
Using Eq. \eqref{Nmod2} in Eq. \eqref{Hkenergy} one obtains
\be\label{Hk3}
\ln 
H_{k}=\ln\left(\frac{k}{a_{0}}\right)+\ln\left(1+Z_{\text{eq(MR)}}
\right)-\ln\rho_{\text{eq(MR)}}^{1/4}+\frac{1}{4}N_{\text{mod}}+
\frac{3\bar{w}_{\text{reh}-1}}{12(1+\bar{w}_{\text{reh}})}\ln\rho_{\text{reh}}
\nn\\ +\frac{1}{3(1+\bar{w}_{\text{reh}}) }\ln\rho_{\text{end}}+\Delta N_{k}.
\ee
From Eq. \eqref{Hk3} we can write 
\be\label{ch9eq6}
-\frac{1}{4}N_{\text{mod}}=\ln\left(\frac{k}{a_{0}}\right)+\ln(1+Z_{\text 
{eq(MR)}})-\ln\rho_{\text{eq(MR)}}^{1/4}+\frac{3\bar{w}_{\text{reh}}-1}{
12(1+\bar {
w}_{\text{reh}})}\ln\rho_{\text{reh}}\nn\\
+\frac{1}{3(1+\bar{w}_{\text{reh}})}
\ln\rho_{\text{end}}+\Delta N_{k}-\ln H_{k}.
\ee
Now, the $N_{\text{mod}}$ can be expressed in terms of modulus mass, 
$m_{\chi}$, and lifetime of the modulus by computing the evolution of the scale 
factor from the time of modulus radiation equality $t_{\text{eq}}$ to time of 
modulus decay $t_{decay}$. The scale factor at any time between the modulus 
radiation equality and modulus decay can be written as 
\be
a(t)=a_{\text{eq(mod)}}\left(\frac{3}{2}H_{\text{eq(mod)}}\left(t-t_{
\text{eq(mod)}}\right)+1\right)^{\frac{2}{3}},
\ee
which gives the number of e-folds during modulus dominated era as 
\be\label{Nmoddom1}
N_{\text{mod}}&=&\int{H~dt}=\int_{t_{\text{eq(mod)}}}^{t_{\text{decay}}}{H_{
\text { eq(mod) } } \left(\frac{3}{2}H_{\text{eq(mod)}}\left(t-t_{
\text{eq(mod)}}\right)+1\right)^{-1}~dt}\nn\\
&=& \frac{2}{3}\ln \left(\frac{3}{2}H_{\text{eq(mod)}}\left(t-t_{
\text{eq(mod)}}\right)+1\right)\nn\\
\ee
If we consider that the lifetime of modulus $\tau_{\text{mod}}$ is the time 
elapsed between $t_{\text{end}}$ and $t_{\text{decay}}$ then one can write 
\be\label{HTau}
H_{\text{eq(mod)}}\left(t_{\text{decay}}-t_{\text{eq(mod)}}\right)
&=&H_{\text{eq(mod)}}\left(t_{\text{decay}}-t_{\text{end}}\right)
-H_{\text{eq(mod)}}\left(t_{\text{eq(mod)}}-t_{\text{reh}}\right)\nn\\
&&\qquad - H_{\text{eq(mod)}}\left(t_{\text{reh}}-t_{\text{end}}\right)\nn\\
&=&H_{\text{eq(mod)}}\tau_{\text{mod}}-\frac{H_{\text{eq(mod)}}}{2 
H_{\text{reh}}}\left(\frac{a_{\text{eq(mod)}}}{a_{\text{reh}}}\right)^{2}\nn\\
&&\qquad -\frac{2}{3(1+\bar{w}_{\text{reh}})}\frac{H_{\text{eq(mod)}}}{H_{\text{end}}}
\left(\frac{a_{\text{reh}}}{a_{\text{end}}}\right)^{\frac{3(1+\bar{w}_{\text{reh
} } ) } { 2 } }\nn\\\
&=&H_{\text{eq(mod)}}\tau_{\text{mod}}-\frac{1}{2}-\frac{2}{3(1+\bar{w}_{\text{
reh } } ) }e^{-2N_{\text{rad}}}
\ee
Substituting Eq. \eqref{HTau} in Eq. \eqref{Nmoddom1} we obtain 
\be\label{Nmoddom2}
N_{\text{mod}}&=&\frac{2}{3}\ln 
\left(\frac{3}{2}H_{\text{eq(mod)}}\tau_{\text{mod}}-\frac{3}{4}-\frac{1}{1+\bar
{w}_{\text{reh}}}e^{-2N_{\text{rad}}} +1\right)\nn\\
&\approx& \frac{2}{3}\ln \frac{3}{2}+\frac{2}{3}\ln 
\left(H_{\text{eq(mod)}}\tau_{\text{mod}}\right) 
\ee
Now, employing Eq. \eqref{ch9rhoeq} we can compute $H_{\text{eq(mod)}}$, and 
substituting it in Eq. \eqref{Nmoddom2} we obtain 
\be\label{ch9eq7}
N_{\text{mod}}\approx -\frac{2}{3}\ln 3-\frac{5}{3} \ln 2
+\frac{2}{3} \ln m_{\chi}\tau +\frac{8}{3}\ln Y
\ee

where $m_{\chi}$ is the modulus mass, and the initial displacement of the 
modulus field $\chi$ is defined as $\chi_{\text{in}}=Y M_{\text{Pl}}$.
Now, substituting Eq. \eqref{ch9eq7} in Eq. \eqref{ch9eq6} we get
\be\label{moduli1}
\frac{2}{3}\ln 3+\frac{5}{3} \ln 2-\frac{1}{6}\ln 
(m_{\chi}\tau)-\frac{2}{3}\ln 
Y=\ln\left(\frac{k}{a_{0}}\right)+\ln\left(1+Z_{\text{eq(MR)}}\right)-\ln\rho_{
\text{eq(MR)}}^{1/4}\nn\\
+\frac{3\bar{w}_{\text{reh}}-1}{12(1+\bar{w}_{\text{reh}})}
\ln\rho_{\text{reh}}-\frac{1}{3(1+\bar{w}_{\text{reh}})}\ln\rho_{\text{end}}
+\Delta N_{k}-\ln H_{k}.
\ee
To make a contact with the slow-roll inflation, we begin by the definition 
of the slow-roll parameter, $\epsilon$, as
\be\label{epsilon}
\epsilon=-\frac{\dot{H}}{H^{2}}=\frac{\frac{3}{2}\dot{\phi}^{2}}{\frac{1}{2}\dot
{\phi}^{2}+V(\phi)}.
\ee
From Eq. \eqref{epsilon}, the kinetic energy of the inflaton field can be 
expressed in terms of the slow-roll parameter $\epsilon$ as 
\be
\frac{1}{2}\dot{\phi}^{2}=\frac{\epsilon V(\phi)}{3-\epsilon}.
\ee
Now, we can write the energy density of the Universe and the Hubble parameter 
during inflation as a function of the slow-roll parameter $\epsilon$ as follows 
\be\label{rho1}
\rho(\phi)=\frac{1}{2}\dot{\phi}^{2}+V(\phi)=\frac{3V(\phi)}{3-\epsilon},
\ee
\be
H^{2}=\frac{\rho}{3M_{\text{Pl}}^{2}}=\frac{1}{M_{\text{Pl}}^{2}}
\left(\frac{V(\phi)}{3-\epsilon}\right).
\ee
At the end of inflation, the slow-roll parameter becomes of the order of unity, 
$\epsilon\sim 1$. Hence, the energy density of the Universe at the end of 
inflation is $\rho_{\text{end}}=\frac{3}{2} V_{\text{end}}$, with 
$V_{\text{end}}$ being the potential at the end of inflation. 
Therefore, employing Eq. \eqref{rho1} in Eq. \eqref{moduli1}, the modulus mass 
can be expressed in terms of the inflationary and reheating parameters as given 
below 
\be\label{moduli2}
m_{\chi}&\approx 4\sqrt{\pi}M_{\text{Pl}}\text{Exp}\Bigg\{-\Big(\frac{2}{3}\ln 
3+\frac{5}{3} \ln 2
-\frac{2}{3}\ln 
Y-\ln\left(\frac{k}{a_{0}}\right)-\ln(1+Z_{\text{eq(MR)}})+\ln\rho_{\text{eq 
(MR)}}^{
1/4}\nn\\
&-\frac{3\bar{w}_{\text{reh}}-1}{12(1+\bar{w}_{\text{reh}})}\ln\left(\frac{\pi^{
2 } } { 30 }g_{\text{reh}}T^{4}_{\text{reh}}\right)
-\frac{1}{3(1+\bar{w}_{\text{reh}})}\ln\left(\frac{3}{2}V_{\text{end}}
\right)-\Delta N_{k}+\ln H_{k}\Big)\Bigg\}
\ee
Eq. \eqref{moduli2} is the key relationship we shall use for relating the 
moduli mass, late time observables and reheating parameters for different 
inflationary models. We consider $Y=1/10$ as per Ref.s \cite{Dutta:2014tya, 
Das:2015uwa, Allahverdi:2018iod}, and  $g_{\text{reh}}\sim 100$ \cite{Dai:2014jja} for our calculations. 

Before we commence our analysis, we point out that in ref. \cite{Dutta:2014tya}
the parameter $w_{\text{re}}$ is quite reasonably assumed to be $w_{\text{re}}<\frac{1}{3}$. Ref  \cite{Das:2015uwa}
introduces two parameters $w_{\text{re1}}$ and $w_{\text{re2}}$ for the two reheating phases
and the same assumption on the limits to their values is made. We are able to reproduce the bound 
$m_{\chi }\gtrsim 10^9$GeV as obtained in these works, when we use the simpler
parameter values  used there, though we use the updated data of Planck 2018. However, due to the exponential nature
of some of the dependences, it becomes important to study the variation due to $\bar{w}_{\text{reh}}$, and the limited range
of its allowed values then sharpens the constraints substantially as will be seen in the following.

It is also to be noted that Planck 2018 results \cite{Akrami:2018odb} study model dependence of the constraints on the 
tensor to scalar ratio $r$.  Of the several parameters whose effects are studied, there is a significant shift  in 
the $n_s$ range when $N_{\text{eff}}$ is varied. This would affect several of our answers 
for $m_\chi$ by a few orders of magnitude, indeed providing physically accepted values in models where 
such would have been completely excluded by the baseline 
$\Lambda$CDM model\footnote{We thank the referee for the suggestion to consider this point}. 
We summarise this possibility in subsection \ref{sec:includeNeff}

\section{Inflationary models and constraints on the lightest modulus 
mass}\label{sec3}
\subsection*{Quadratic large field model:}
\begin{figure}[H]
\begin{subfigure}{.5\textwidth}
 \centering
  \includegraphics[width=.8\linewidth]{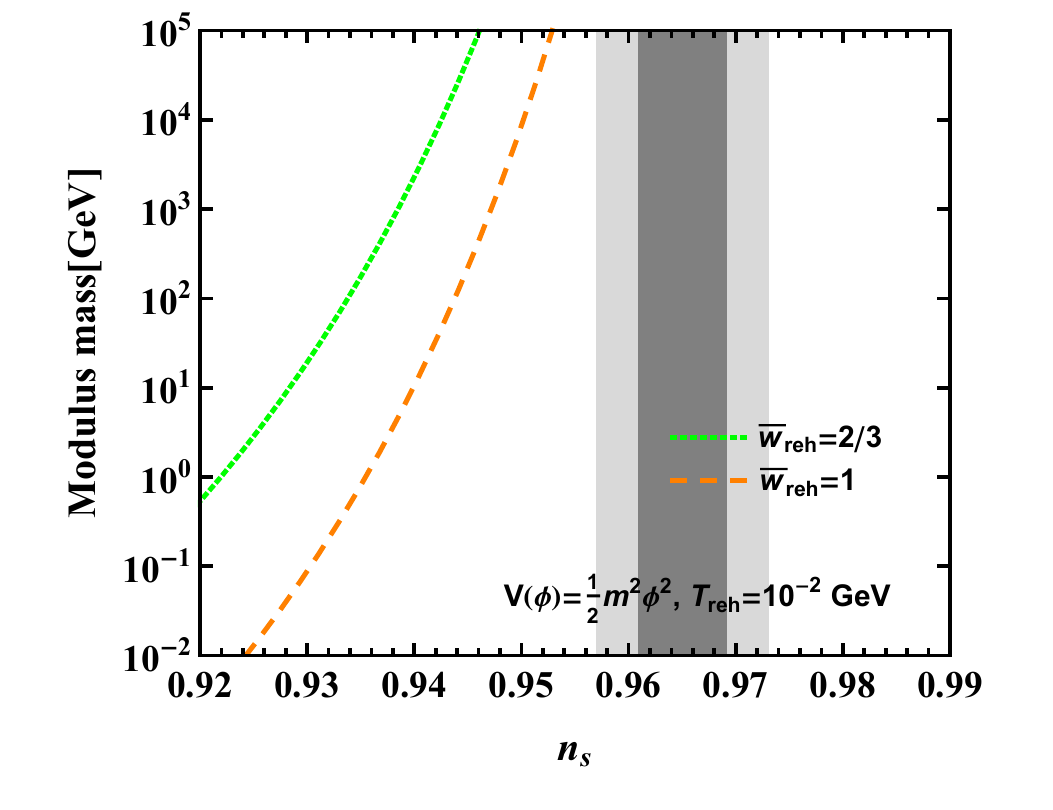}
   \caption{}
   \label{fig1a}
 \end{subfigure}%
 \begin{subfigure}{.5\textwidth}
   \centering
   \includegraphics[width=.8\linewidth]{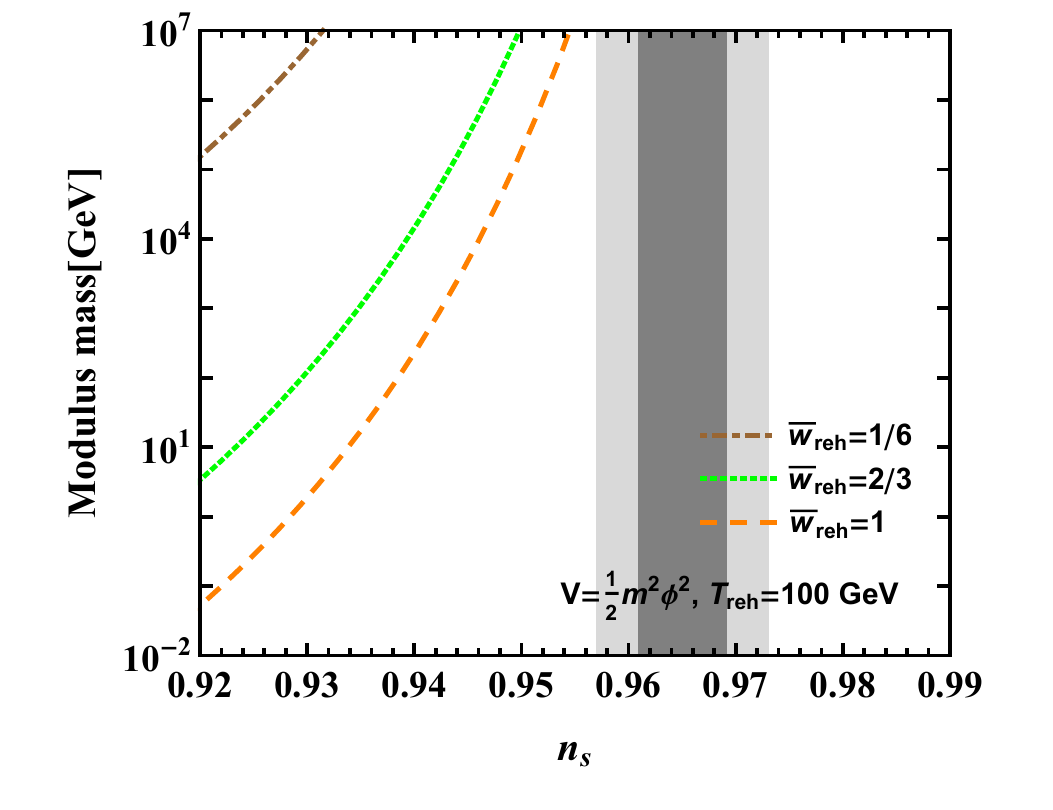}
   \caption{}
   \label{fig1b}
 \end{subfigure}
 \begin{subfigure}{.5\textwidth}
  \centering
   \includegraphics[width=.8\linewidth]{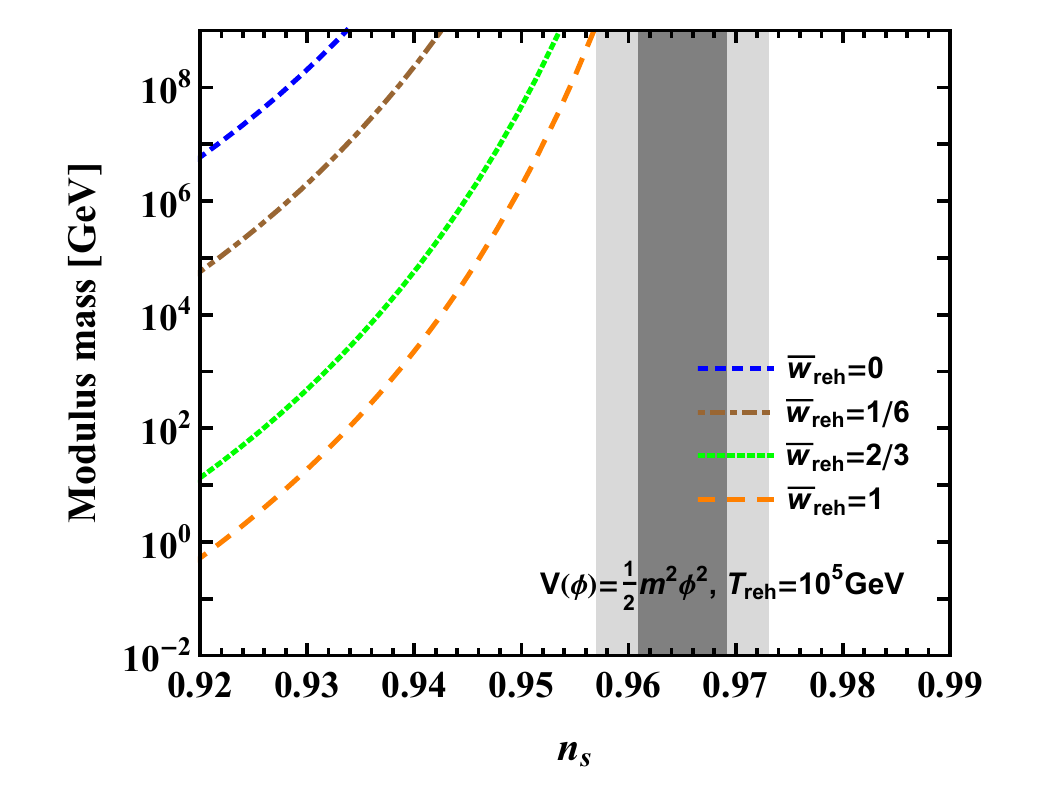}
   \caption{}
   \label{fig1c}
 \end{subfigure}%
 \begin{subfigure}{.5\textwidth}
   \centering
   \includegraphics[width=.8\linewidth]{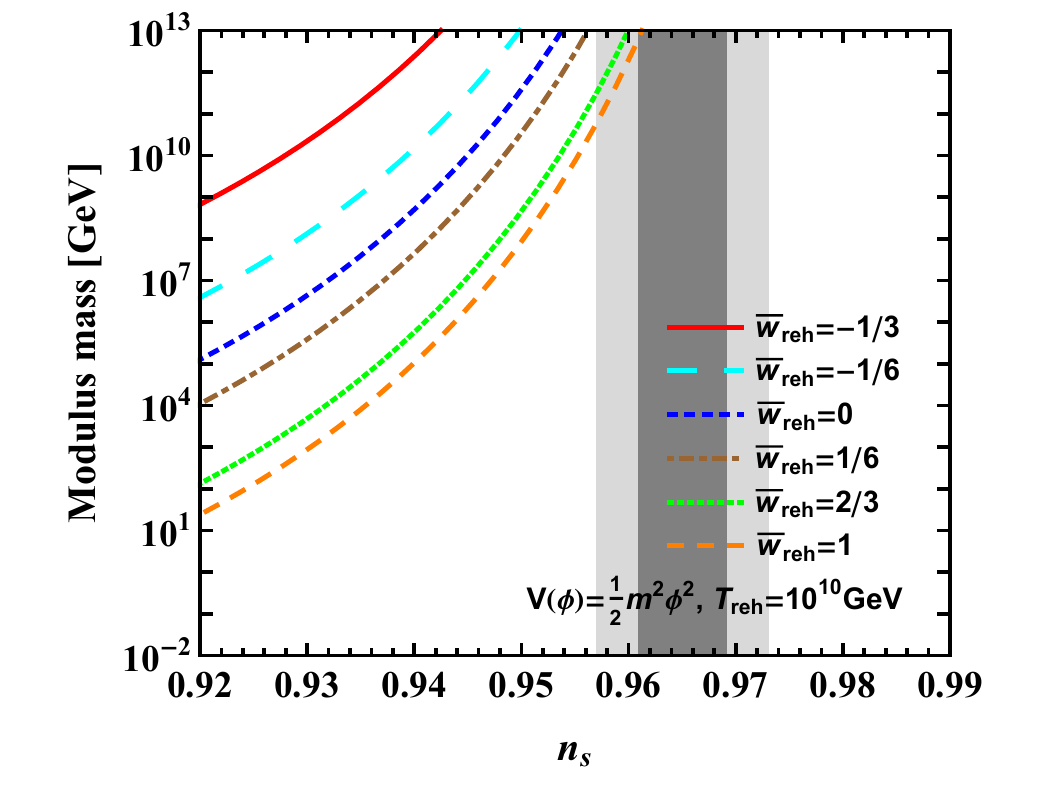}
   \caption{}
   \label{fig1d}
 \end{subfigure}
 \begin{subfigure}{.5\textwidth}
  \centering
   \includegraphics[width=.8\linewidth]{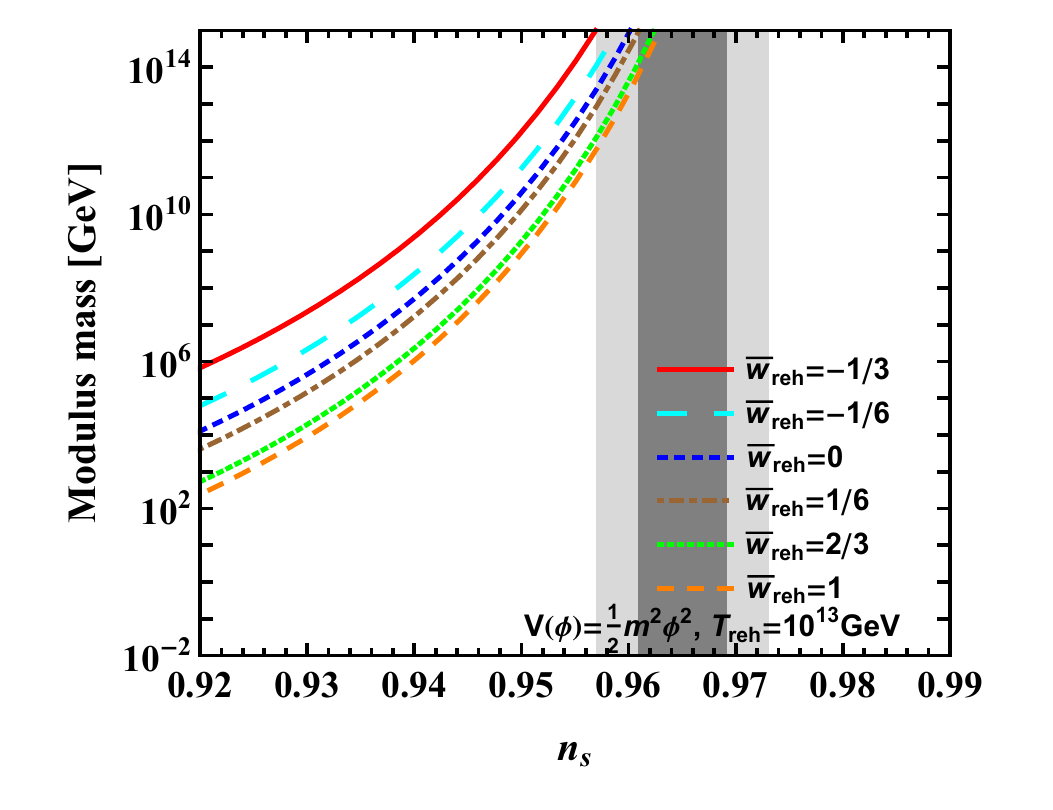}
   \caption{}
   \label{fig1e}
 \end{subfigure}%
 \begin{subfigure}{.5\textwidth}
  \centering
   \includegraphics[width=.8\linewidth]{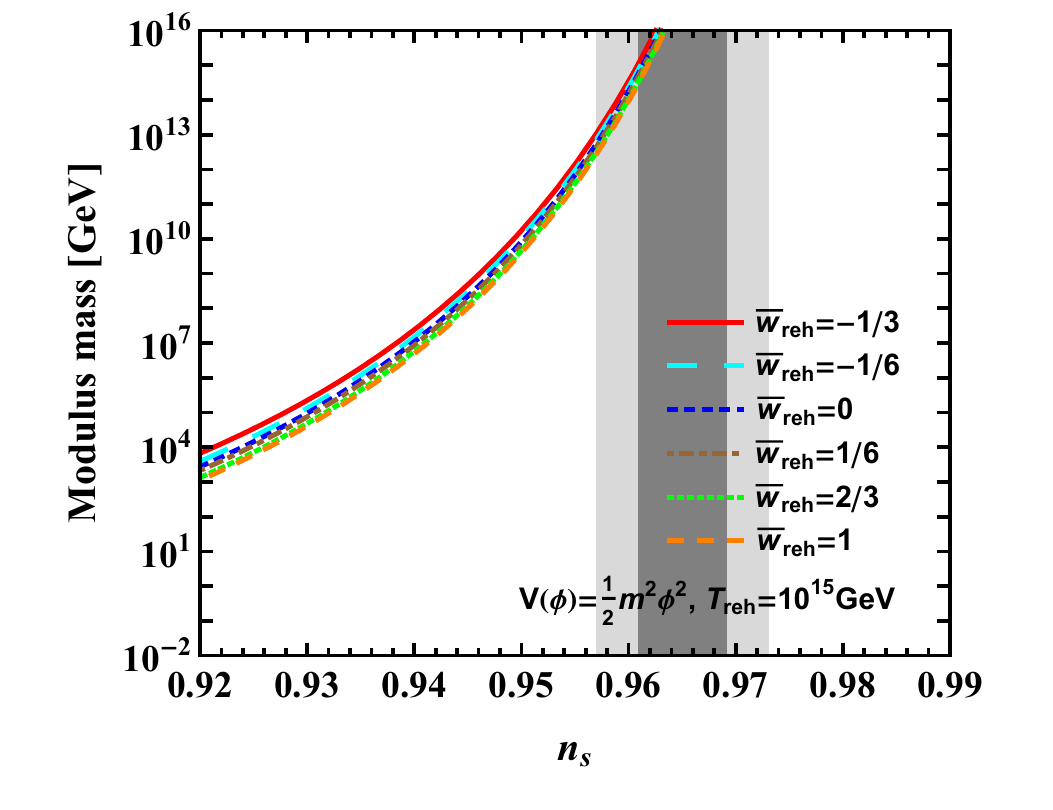}
   \caption{}
   \label{fig1f}
 \end{subfigure}
 \caption{Plots of allowed modulus mass values $m_{\chi}$ as a 
function of $n_{s}$ for the quadratic large field model for different values of 
 $\bar{w}_{\text{reh}}$: 
 $\bar{w}_{\text{reh}}=-\frac{1}{3}$ (solid red), 
 $\bar{w}_{\text{reh}}=-\frac{1}{6}$ (large dashed cyan), 
 $\bar{w}_{\text{reh}}=0$ (small dashed blue), 
 $\bar{w}_{\text{reh}}=\frac{1}{6}$ 
 (dot-dashed brown), 
 $\bar{w}_{\text{reh}}=\frac{2}{3}$ (tiny dashed green), 
 $\bar{w}_{\text{reh}}=1$ 
 (medium dashed orange). The dark gray and light gray shaded 
 regions
 correspond to the 1$\sigma $ and 2$\sigma$ bounds respectively on $n_{s}$ 
from 
 Planck 2018 data (TT, TE, EE + lowE + lensing) \cite{Aghanim:2018eyx}.}
 \label{ch9fig1}
 \end{figure}

The quadratic large field model 
\cite{Linde:1984st,Bassett:2005xm,Martin:2013tda,Dai:2014jja} of 
inflation is 
described by the potential 
$V(\phi)=\frac{1}{2}m^{2}\phi^{2}$. Now, consider the mode $k_{*}$ corresponding 
to the pivot scale introduced above, Eq. \eqref{e7}, which 
crosses the Hubble radius $H_{*}$ during inflation when the field $\phi$ has 
attained the value 
$\phi_{*}$. The number of e-folds remaining after the pivot scale $k_{*}$ 
crosses the Hubble radius is
\be\label{e29}
\begin{aligned}
 \Delta 
N_{*}&\simeq\frac{1}{M_{\text{Pl}}^{2}}\int_{\phi_{\text{end}}}^{\phi_{*}}{\frac
{V}{V'}d\phi}
=\frac{1}{4}\Big{[}\left(\frac{\phi_{*}}{M_{\text{Pl}}}\right)^{2}-2\Big{]},
 \end{aligned}
\ee
where we have used  the condition defining the end of inflation, $\epsilon=1$ 
which gives 
$\frac{\phi_{\text{end}}^{2}}{M_{\text{Pl}}^{2}}=2$.
Using $\epsilon=2M^2_{\text{Pl}}/\phi^{2}$ as arises in this model, the 
spectral index $n_{s}$, Eq. 
\eqref{ns}, can be written as
\be\label{e32}
 n_{s}=1-8\left(\frac{M_{\text{Pl}}}{\phi_{*}}\right)^{2}.
\ee
And thus $\Delta 
N_{*}$ as a function of the scalar spectral index $n_{s}$ and is given by
\be\label{eq:deltaNstarns}
 \Delta N_{*}=\left(\frac{2}{1-n_{s}}-\frac{1}{2}\right).
\ee
Further, in this model one obtains the relation 
\be\label{e36}
 H_{*}=\pi M_{\text{Pl}}\sqrt{2A_{s}(1-n_{s})}
\ee
where $n_s$ although strictly $k$ dependent has been replaced by it almost 
constant value. This, 
along with the relation of $H$ and field $\phi$ in this model, and the criterion 
for the end of 
inflation as used in \eqref{e29}, gives the value of $V$ at
the end 
of the inflation, $V_{\text{end}}$, as a function of $A_{s}$ 
and $n_s$,
\be\label{e37}
V_{\text{end}}=\frac{1}{2}m^{2}\phi_{\text{end}}^{2}
\approx \frac{3}{2}\pi^{2}A_{s}M_{\text{Pl}}^{4}(1-n_{s})^{2} .
\ee
Substituting Eqs. \eqref{e32}, \eqref{e36} and \eqref{e37} in Eq. 
\eqref{moduli2} the modulus mass can be expressed as a function of $n_{s}$ as
\be\label{eq:mchinsquarticlarge}
m_{\chi}&\approx 4\sqrt{\pi}M_{\text{Pl}}\text{Exp}\Bigg\{-\Big(\frac{2}{3}\ln 
3+\frac{5}{3} \ln 2-\frac{2}{3}\ln 
Y-\ln\left(\frac{k_{*}}{a_{0}}\right)-\ln(1+Z_{\text{eq(MR)}})+\ln\rho_{\text{
eq 
(MR)}}^{1/4}\nn\\
&-\frac{3\bar{w}_{\text{reh}}-1}{12(1+\bar{w}_{\text{reh}})}\ln\left(\frac{\pi^{
2}}{30}g_{\text{reh}}T_{\text{reh}}^{4}\right)
-\frac{1}{3(1+\bar{w}_{\text{reh}})}\ln\left(\frac{3}{2}{\pi^{2}A_{s}M_{\text{
Pl}}^{4}(1-n_{s})^{2}}
\right)\nn\\
&-\left(\frac{2}{1-n_{s}}-\frac{1}{2}\right)+\ln\left(\pi
M_{\text{Pl}}\sqrt{2A_{s}(1-n_{s})}\right)\Big)\Bigg\}
\ee
The variation of the modulus mass, $m_{\chi}$, as a function of the scalar 
spectral index $n_{s}$ for different values of 
$\bar{w}_{\text{reh}}$ and reheating temperature $T_{\text{reh}}$ are shown in 
figure \ref{ch9fig1}.  Planck's central value of $A_{s}=2.1\times 10^{-9}$  
and $z_{\text{eq (MR)}}=3402$  are used, and the parameter 
$\rho_{\text{eq (MR)}}$ is computed to be $10^{-9}$ GeV
\cite{Aghanim:2018eyx} to obtain the figure~{\ref{ch9fig1}}. Here, the upper limit of $m_{\chi}$ is taken to be  $m_{\chi}\lesssim \left(T_{\text{reh}}^{2}M_{\text{Pl}}\right)^{\frac{1}{3}}$ as the modulus domination and decay occur after the complition of inflationary reheating, hence, the temperature after the modulus decay $(T_{\text{decay}})$ must be less than $T_{\text{reh}}$. From figure~{\ref{ch9fig1}} we see that for reheating temperature $T_{\text{reh}}<10^{5}$ GeV, curves with $\bar{w}_{\text{reh}}<0$ do not give the modulus mass in the allowed range. For reheating temperature $T_{\text{reh}}<10^{10}$ GeV, within Planck's $1\sigma$ bounds on $n_{s}$, curves with $\bar{w}\gtrsim \frac{1}{6}$ predict the modulus mass in the reasonable range.  However, for $T_{\text{reh}}\ge 10^{10}$ GeV all curves tend to fall in the allowed range of $m_{\chi}$,  and at a temperature around $T_{\text{reh}}\sim 10^{15}$ 
GeV all curves converge within Planck's $1\sigma$ bound on $n_{s}$, which corresponds to an instantaneous reheating (see 
Ref. \cite {Goswami:2018vtp}).
 \subsection*{Quartic hilltop potential:}
\begin{figure}[H]
\begin{subfigure}{.5\textwidth}
 \centering
  \includegraphics[width=.8\linewidth]{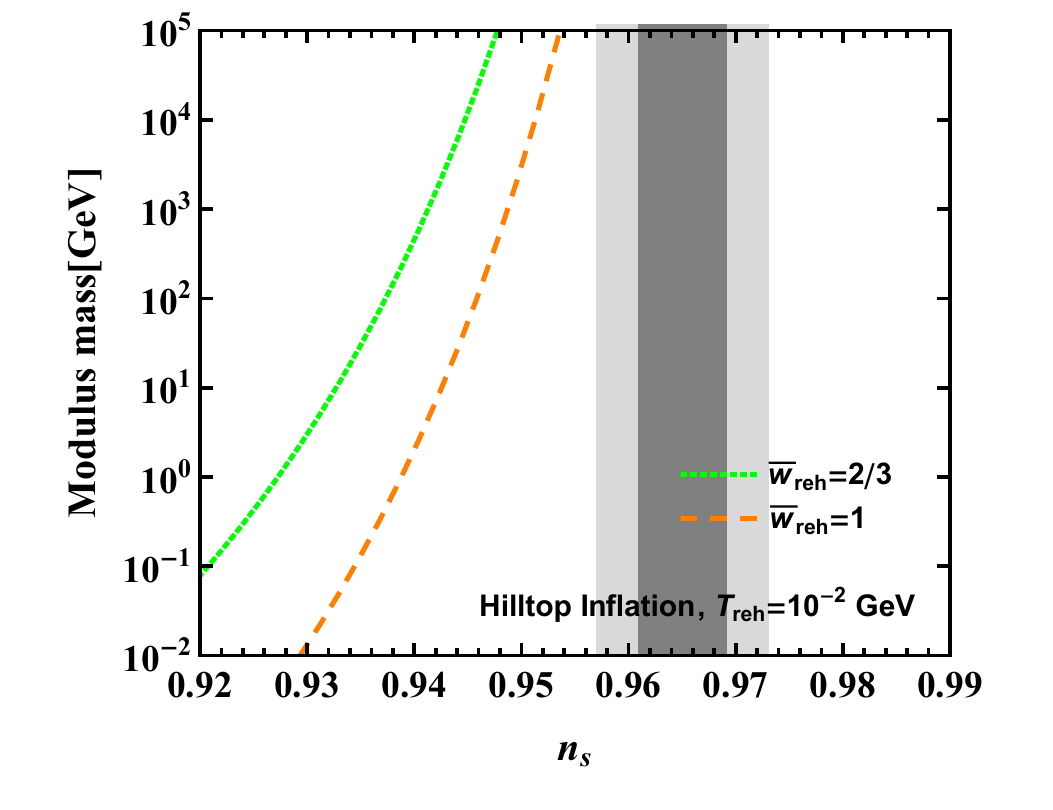}
  \caption{}
  \label{fig2a}
\end{subfigure}%
\begin{subfigure}{.5\textwidth}
  \centering
  \includegraphics[width=.8\linewidth]{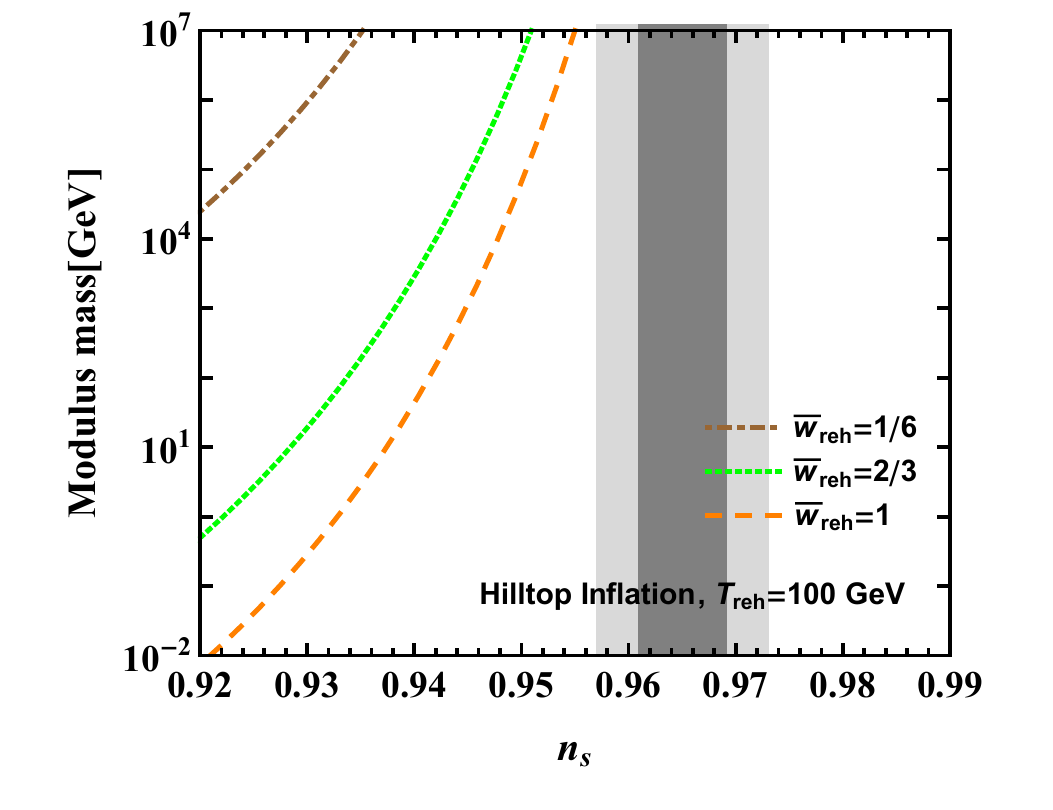}
  \caption{}
  \label{fig2b}
\end{subfigure}
\begin{subfigure}{.5\textwidth}
 \centering
  \includegraphics[width=.8\linewidth]{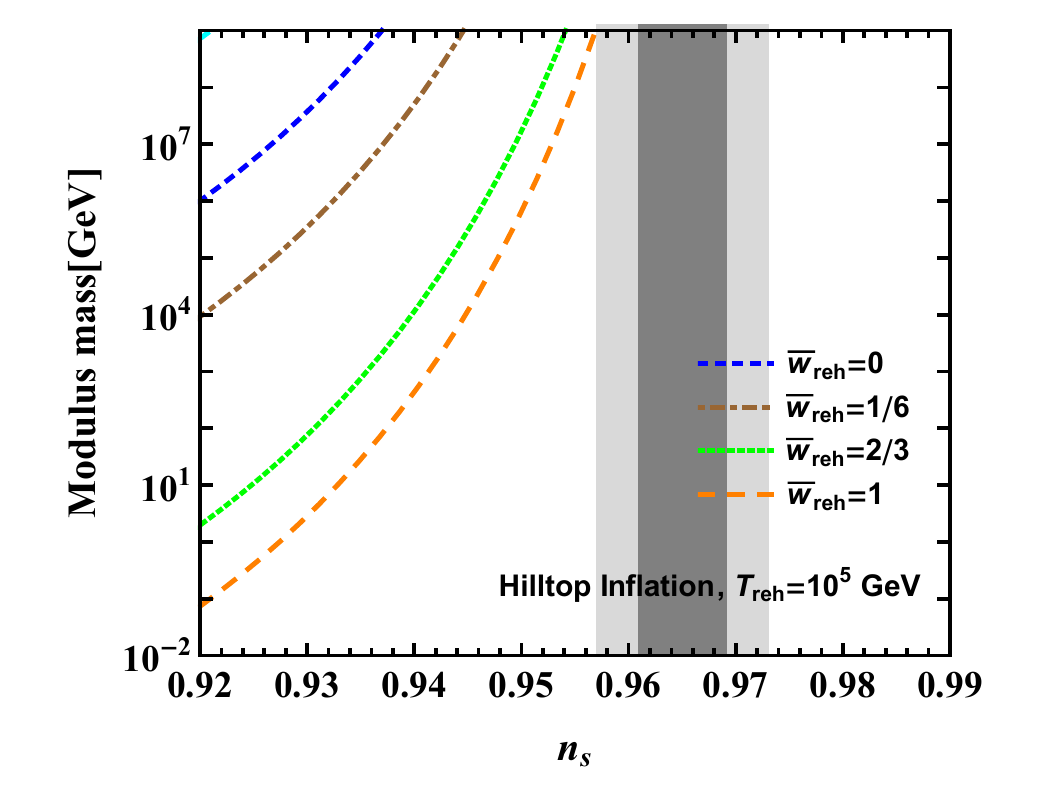}
  \caption{}
  \label{fig2c}
\end{subfigure}%
\begin{subfigure}{.5\textwidth}
  \centering
  \includegraphics[width=.8\linewidth]{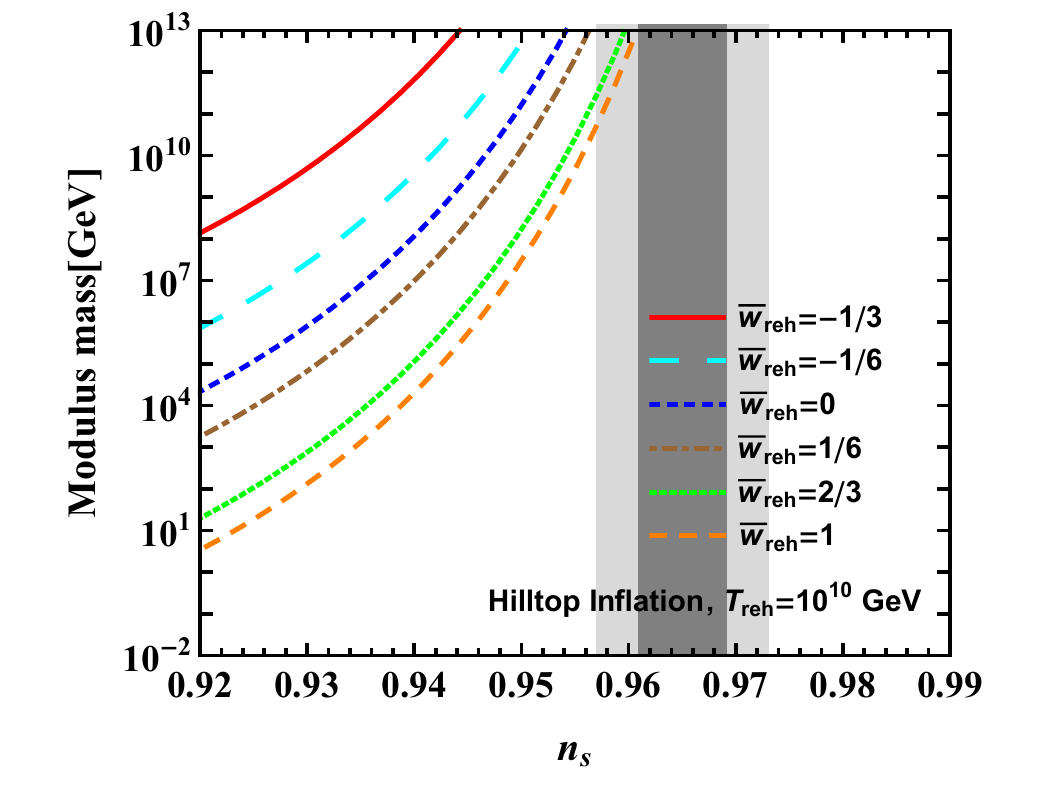}
  \caption{}
  \label{fig2d}
\end{subfigure}
\begin{subfigure}{.5\textwidth}
 \centering
  \includegraphics[width=.8\linewidth]{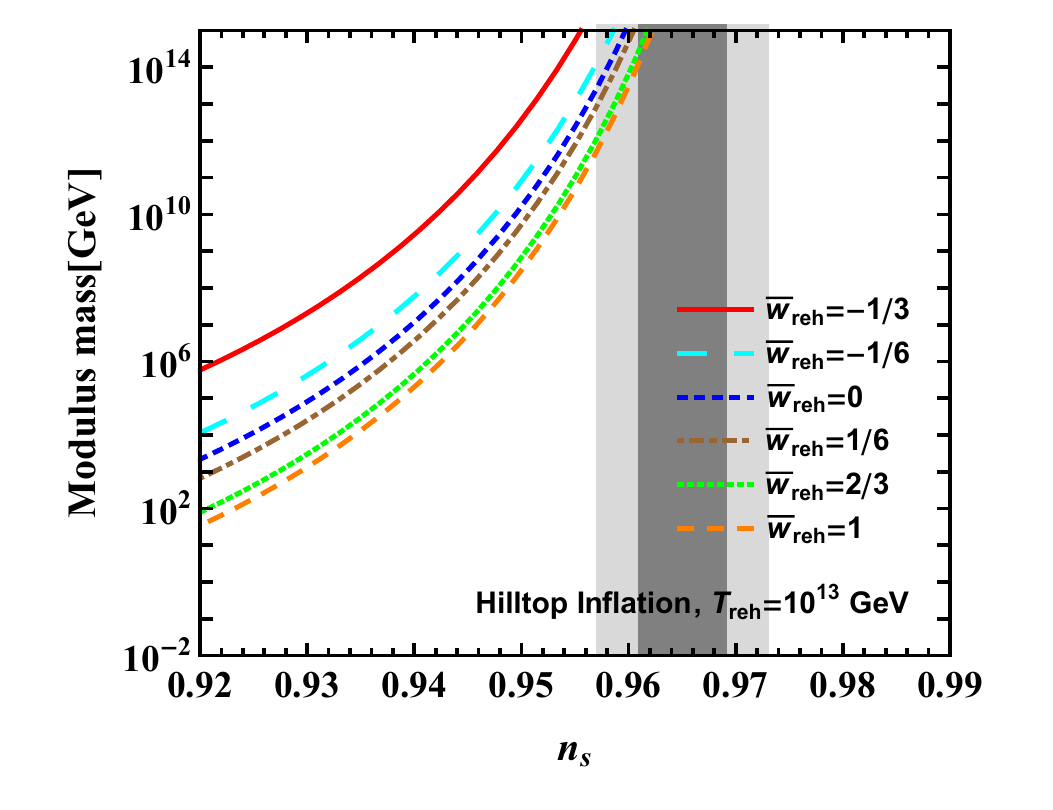}
  \caption{}
  \label{fig2e}
\end{subfigure}%
\begin{subfigure}{.5\textwidth}
 \centering
  \includegraphics[width=.8\linewidth]{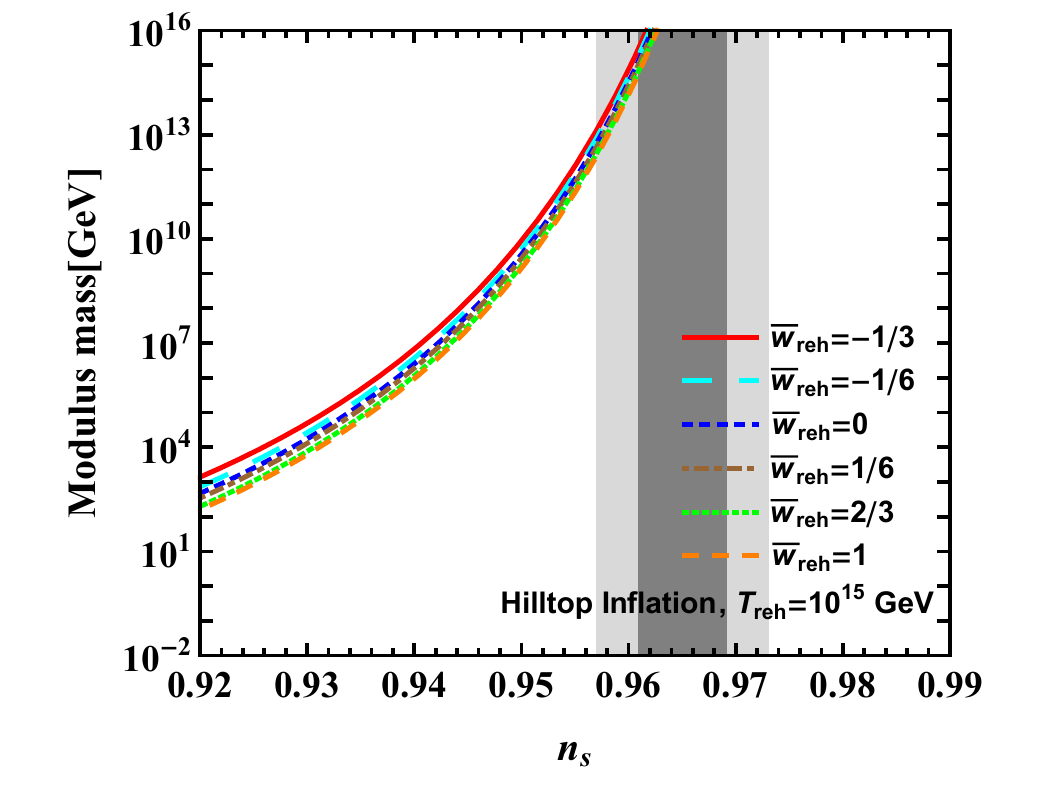}
  \caption{}
  \label{fig2f}
\end{subfigure}
\caption{Plots of allowed modulus mass values $m_{\chi}$ as a 
function of $n_{s}$ for the quartic hilltop potential. All curves and shaded 
regions are as for figure \ref{ch9fig1}.}
\label{ch9fig2}
\end{figure}

In this model, inflation occurs at very small value of the field and at the top 
of the flat 
potential. 
 The potential for this kind of inflation is described by 
\cite{Linde:1981mu,Kinney:1995ki,Martin:2013tda}.
\be\label{small}
V(\phi)= V_{0}\left[1-\left(\frac{\phi}{\mu}\right)^{p}\right] .
\ee
The field value at the end of inflation is calculated by setting $\epsilon=1$ 
and $\phi_{\text{end}}< \mu$ 
which leads 
to the following equation
\be 
\left(\frac{\phi_{\text{end}}}{\mu}\right)^{p}+\frac{p}{\sqrt{2}}\frac{M_{\text{
Pl}}}{\mu}\left(\frac{\phi_{
\text{end}}}{\mu}\right)^{p-
1}=1 .
\ee
As per Ref. \cite{Goswami:2018vtp, Hazra:2010ve}, we have considered $p=4$ and 
$\mu=15 
M_{\text{Pl}}$ and obtained 
$\frac{\phi_{\text{end}}}{M_{\text{Pl}}}=14.34$. For this quartic hilltop model 
we obtain 
\be\label{e46prime}
 \Delta 
N_{*}=6.328\times10^{3}\left[\left(\frac{M_{\text{Pl}}}{\phi_{*}}\right)^{2}
-\left(\frac{1}{14.34}\right)^{2}
\right]
+\frac{1}{8}\left[\left(\frac{\phi_{*}}{M_{\text{Pl}}}\right)^{2}-(14.34)^{2}
\right] .
\ee
We can write the field value at the time of horizon crossing of the pivot scale 
as a function of $n_{s}$ as given below
\be\label{e47}
\begin{aligned}
 n_{s}&=1-6\epsilon_{*}+2\eta_{*}\\
&=1-3M_{\text{Pl}}^{2}\left(-\frac{4\phi_{*}^{3}}{(15M_{\text{Pl}})^{4}}-\phi_{*
}^{4}\right)^{2}-\frac{24\phi_ 
{*}^{2}}{(15M_{Pl})^ {4} -\phi_ {*}^{4}}M_{\text{Pl}}^{2}
 \end{aligned}
\ee
The $H_{*}$ and $V_{\text{end}}$ can be expressed as a function of $A_{s}$ and 
$n_{s}$ as
\be
 H_{*}&=&8\pi 
M_{\text{Pl}}\left(\frac{\chi^{3}(n_{s})}{15^{4}-\chi^{4}(n_{s})}\right)\sqrt{A_
{s}}=8\pi 
M_{\text{Pl}}\beta(n_{s})\sqrt{A_{s}},\label{Hsmall}\\
\label{Vend}
 V_{\text{end}}&=&\gamma A_{s} 
M_{\text{Pl}}^{4}\frac{\beta^{3}(n_{s})\left(3-8\beta^{2}(n_{s})\right)}{\chi^{3
}(n_{s})},
\ee
where, $\chi(n_{s})=\frac{\phi_{*}}{M_{\text{Pl}}}(n_{s})$ is the 
solution of equation Eq. \eqref{e47}. Here we define  
 $\beta(n_{s})=\frac{\chi^{3}(n_{s})}{15^{4}-\chi^{4}(n_{s})}$ and  
$\gamma=5.28\times 
10^{6}$. Using the above expressions one can write out $m_{\chi}$ as a function 
of $A_{s}$ and $n_{s}$ for this quartic hilltop potential, and is given below 
\be\label{mchismall}
m_{\chi}&\approx 4\sqrt{\pi}M_{\text{Pl}}\text{Exp}\Bigg\{-\Big(\frac{2}{3}\ln 
3+\frac{5}{3} \ln 2
-\frac{2}{3}\ln 
Y-\ln\left(\frac{k_{*}}{a_{0}}\right)-\ln(1+Z_{\text{eq(MR)}})+\ln\rho_{\text{eq
} } ^ {
1/4}\nn\\
&-\frac{3\bar{w}_{\text{reh}}-1}{12(1+\bar{w}_{\text{reh}})}\ln\left( 
\frac{\pi^{2}}{30}g_{\text{reh}} T_{\text{reh}}^{4}\right)
-\frac{1}{3(1+\bar{w}_{\text{reh}})}\ln\frac{\left(\frac{3}{2}\gamma
A_{s}M_{\text{Pl}}^{4}\beta(n_{s})\left(3-8\beta^{2}(n_{s})\right)
\right)}{\chi^{3}(n_{s})}\nn\\
&-\Delta N_{*}(n_{s})+\ln 
\left(8\pi M_{\text{Pl}}\beta(n_{s})\sqrt{A_{s}}\right)\Big)\Bigg\}.
\ee
Using the above expression, Eq. \eqref{mchismall}, we plot the variation of 
$m_{\chi}$ with $n_{s}$  for six different values of $\bar{w}_{\text{reh}}$ and 
$T_{\text{reh}}$ in figure \ref{ch9fig2}.  Within Planck's $1\sigma$ bounds on 
$n_{s}$, for reheating temperature $T_{\text{reh}}<10^{5}$ GeV, none of the curves predict modulus mass in the allowed reasonable range. 
For higher reheating temperature, i.e., $T_{\text{reh}}>10^{10}$ GeV, it is possible to obtain  
$m_{\chi}$ in the allowed range. However, for $10^{10}$ GeV $\lesssim  $ $T_{\text{reh}} \lesssim 10^{13}$ GeV, curves with $\bar{w}_{\text{reh}}\lesssim 0$ do not predict $m_{\chi}$ within the expected reasonable range within Planck's $1\sigma$ bounds on $n_{s}$. For further increase of $T_{\text{reh}}$ it is possible to obtain the reasonable range of $m_{\chi}$ by considering any allowed value of $\bar{w}_{\text{reh}}$ within Planck's $1\sigma$ bounds on $n_{s}$ 
\subsection*{Starobinsky model:} 
\begin{figure}[H]
\begin{subfigure}{.5\textwidth}
 \centering
  \includegraphics[width=.8\linewidth]{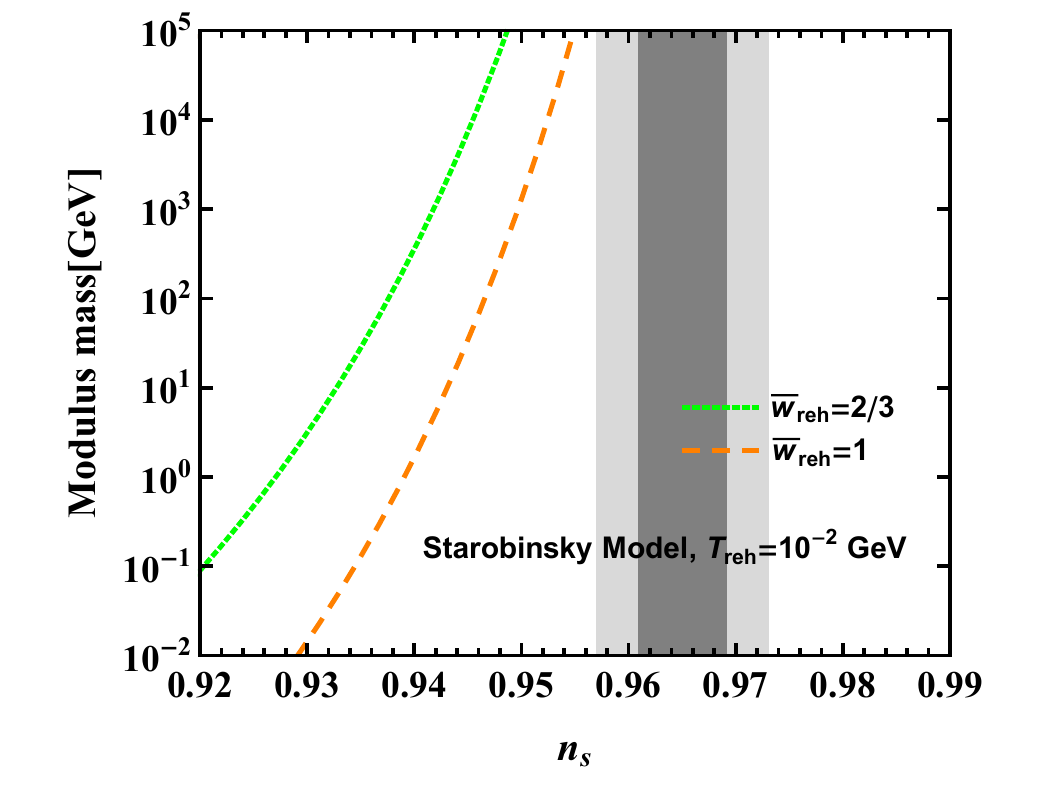}
  \caption{}
  \label{fig3a}
\end{subfigure}%
\begin{subfigure}{.5\textwidth}
  \centering
  \includegraphics[width=.8\linewidth]{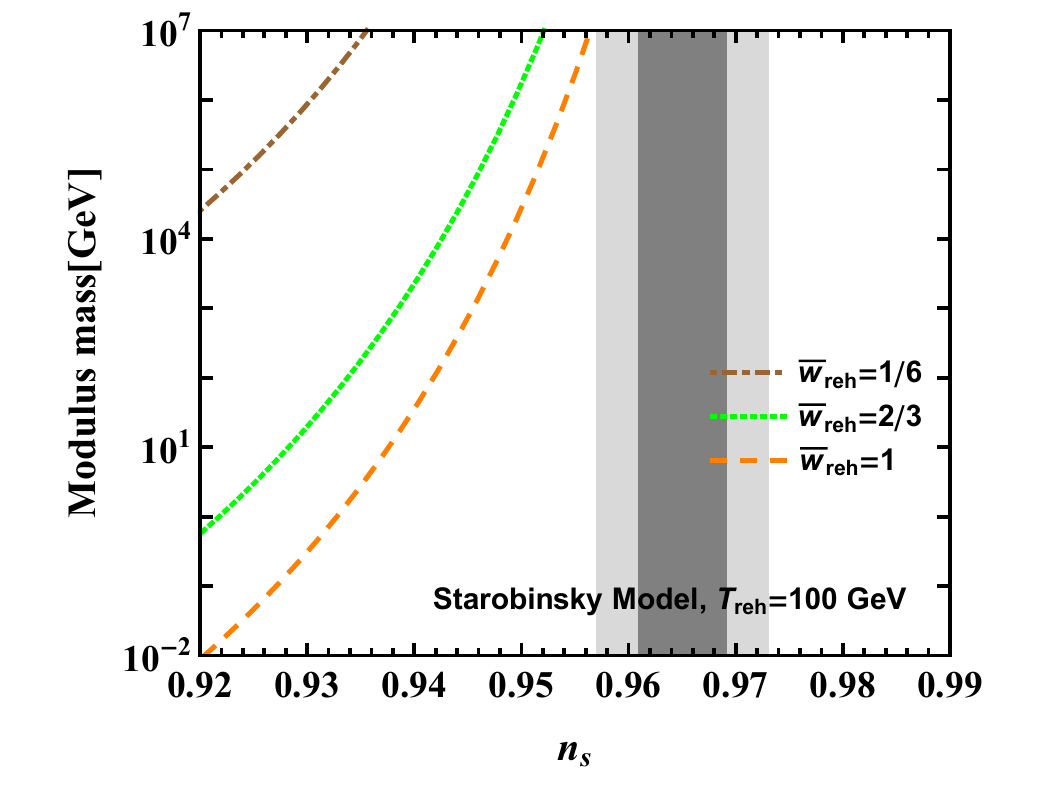}
  \caption{}
  \label{fig3b}
\end{subfigure}
\begin{subfigure}{.5\textwidth}
 \centering
  \includegraphics[width=.8\linewidth]{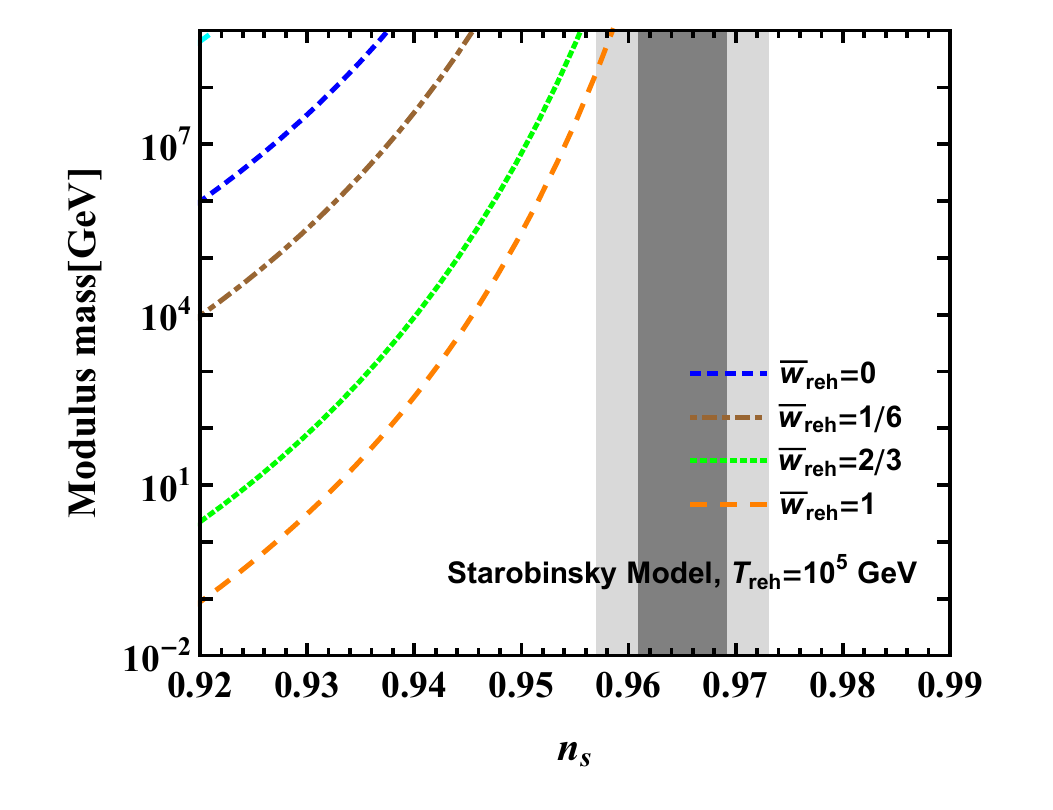}
  \caption{}
  \label{fig3c}
\end{subfigure}%
\begin{subfigure}{.5\textwidth}
  \centering
  \includegraphics[width=.8\linewidth]{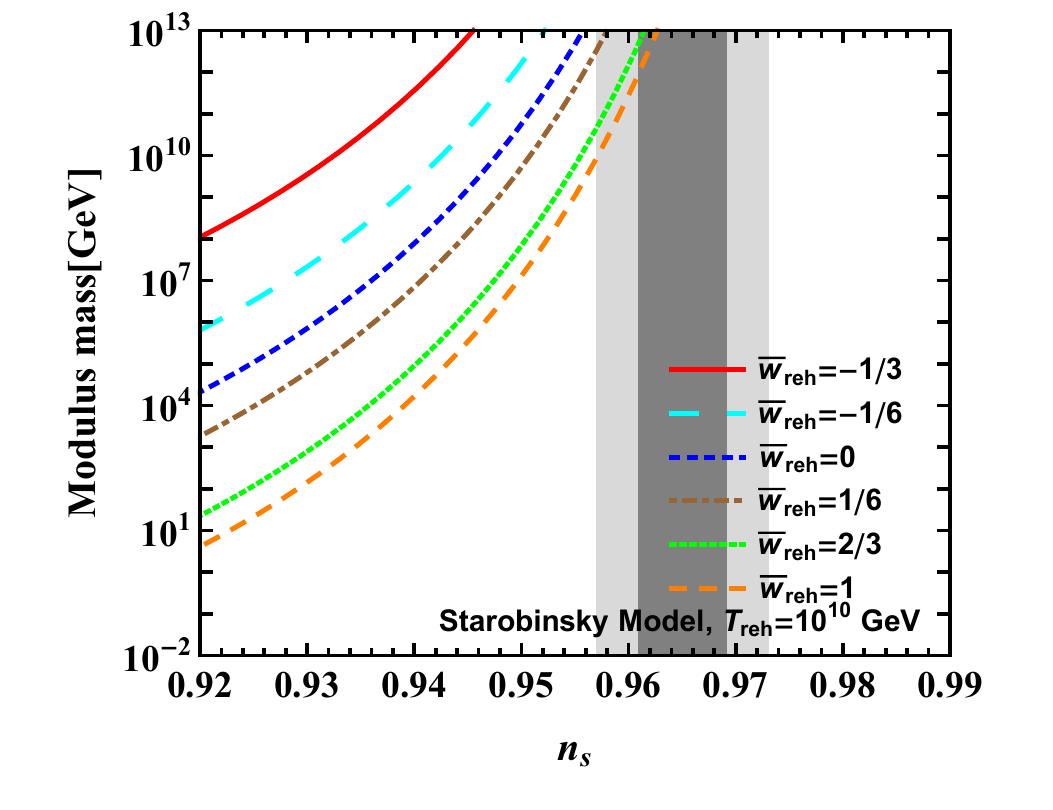}
  \caption{}
  \label{fig3d}
\end{subfigure}
\begin{subfigure}{.5\textwidth}
 \centering
  \includegraphics[width=.8\linewidth]{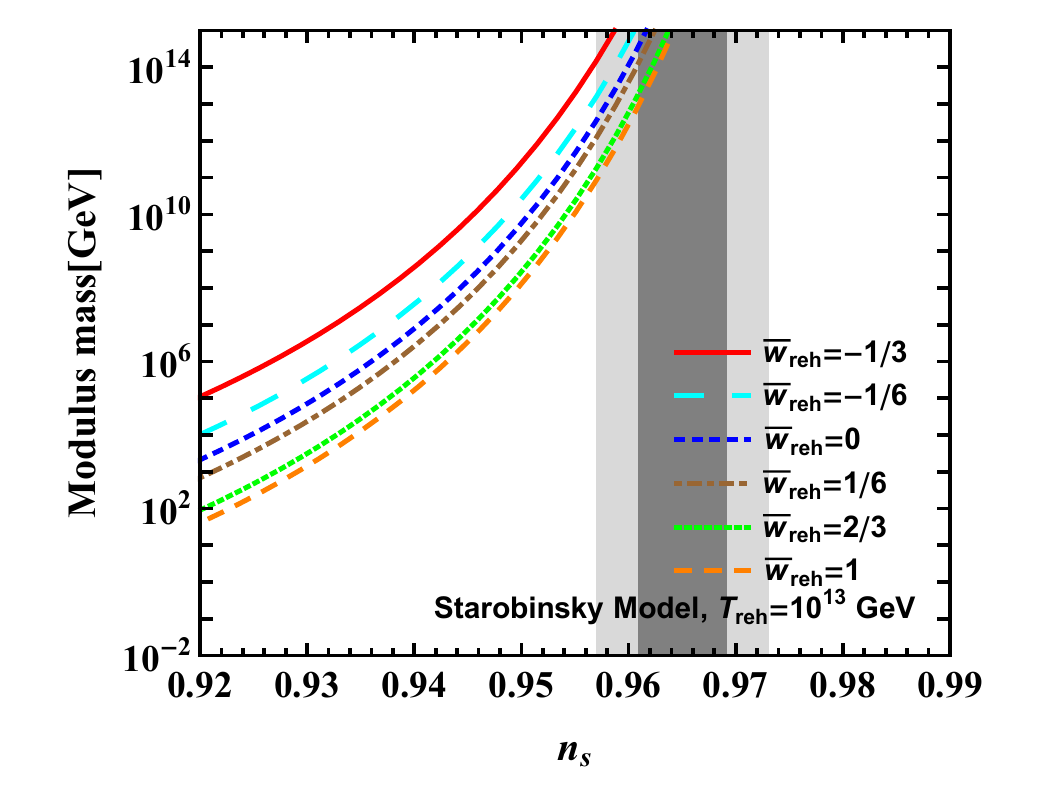}
  \caption{}
  \label{fig3e}
\end{subfigure}%
\begin{subfigure}{.5\textwidth}
 \centering
  \includegraphics[width=.8\linewidth]{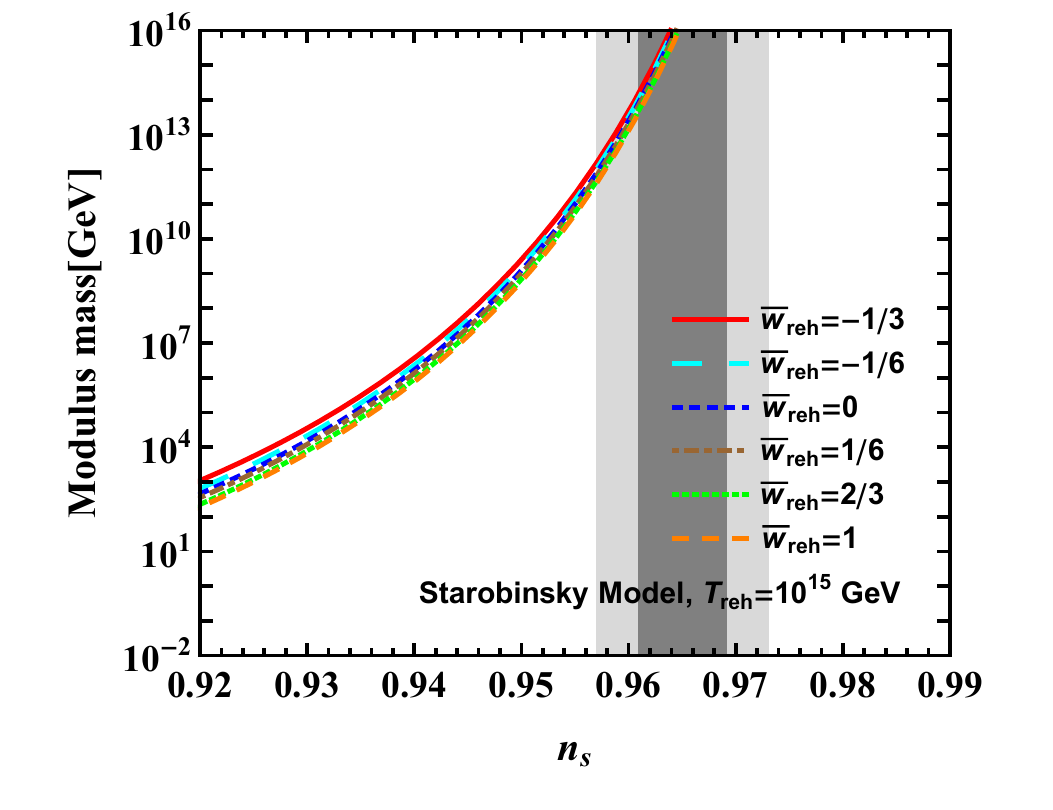}
  \caption{}
  \label{fig3f}
\end{subfigure}
\caption{Plots of allowed modulus mass values $m_{\chi}$ as a 
function of $n_{s}$ for the Starobinsky model. All curves and shaded regions are as 
for figure \ref{ch9fig1}.}
\label{ch9fig3}
\end{figure}

The potential for the Starobinsky model can be written as 
\cite{Kehagias:2013mya,Cook:2015vqa}
\be\label{StrP}
V(\phi)=\Lambda^{4}
\left(1-e^{-\sqrt{\frac { 2 } {3}}\phi/M_{\text{Pl}}} \right)^{2},
\ee
Where $\Lambda$ is the energy scale. Similar to the large field and hilltop 
model, the $\Delta N_{*}$, $H_{*}$ and $V_{\text{end}}$ can be expressed as a 
function of $n_{s}$, and are given below (see Ref. \cite{Goswami:2018vtp} for 
details calculation)
\be
\Delta 
N_{*}=\frac{3}{4}\left[\frac{8}{3(1-n_{s})}-\left(1+\frac{2}{\sqrt{3}}
\right)-\ln\left(\frac{8}{(1-n_{s}
)(3+2\sqrt{3}) } \right)\right],
\ee
\be\label{StrH}
H_{*}\approx\pi M_{\text{Pl}}\left(1-n_{s}\right)\sqrt{\frac{3}{2}A_{s}} 
\ee
and
\be\label{StrV}
V_{\text{end}}\approx\frac{9}{2}\pi^{2}A_{s}M_{\text{Pl}}^{4}(1-n_{s})^{2}\frac{
1}{\left(1+\frac{\sqrt{3}}{2}
\right)^{2}}.
\ee
Using the above expressions one can write $m_{\chi}$ as a function of 
$n_{s}$ and $A_{s}$ for the Starobinsky model, and is given by 
\be\label{mchistr}
m_{\chi}&\approx 4\sqrt{\pi}M_{\text{Pl}}\text{Exp}\Bigg\{-\Bigg(\frac{2}{3}\ln 
3+\frac{5}{3} \ln 2
-\frac{2}{3}\ln 
Y-\ln\left(\frac{k_{*}}{a_{0}}\right)-\ln(1+Z_{\text{eq(MR)}})+\ln\rho_{\text{eq
} } ^ {
1/4}\nn\\
&-\frac{3\bar{w}_{\text{reh}}-1}{12(1+\bar{w}_{\text{reh}})}\ln\left(\frac{\pi^{
2 }
}{30}g_{\text{reh}}T_{\text{reh}}^{4}\right)
-\frac{1}{3(1+\bar{w}_{\text{reh}})}\ln\left(\frac{27}{4}\pi^{2}A_{s
}M_{\text{Pl}}^{4}(1-n_{s})^{2}\frac{1}{\left(1+\frac{\sqrt{3}}{2}\right)^{2}}
\right)\nn\\
&-\frac{3}{4}\left(\frac{8}{3(1-n_{s})}
-\left(1+\frac{2}{\sqrt{3}}\right)-\ln\left(\frac{8}{(1-n_{s})(3+2\sqrt{3})}
\right)\right)+\ln\left(\pi
M_{\text{Pl}}\sqrt{2A_{s}(1-n_{s})}\right)\Bigg)\Bigg\}
\ee
For the 
Starobinsky model, using Eq. \eqref{mchistr}, the relation between  $m_{\chi}$ 
and $n_{s}$ for different values of $\bar{w}_{\text{reh}}$ and 
$T_{\text{reh}}$ are shown in figure \ref{ch9fig3}. Similar to the other two models, this starobinsky model also does not predict the value of $m_{\chi}$ in the reasonable range for $T_{\text{reh}}<10^{10}$ GeV and within Planck's $1\sigma$ bounds on $n_{s}$. However, with the increase of $T_{\text{reh}}$ all corves move towards the central value of $n_{s}$, and for $T_{\text{reh}}\gtrsim 10^{13}$ GeV all curves predict the value of $m_{\chi}$ in the allowed reasonable range within Plancks $2\sigma$ bounds on $n_{s}$.
\section{Features in the inflaton potential and constraints on modulus mass} 
\label{Sec4}
\begin{figure}[H]
 \centering
  \includegraphics[width=.5\linewidth]{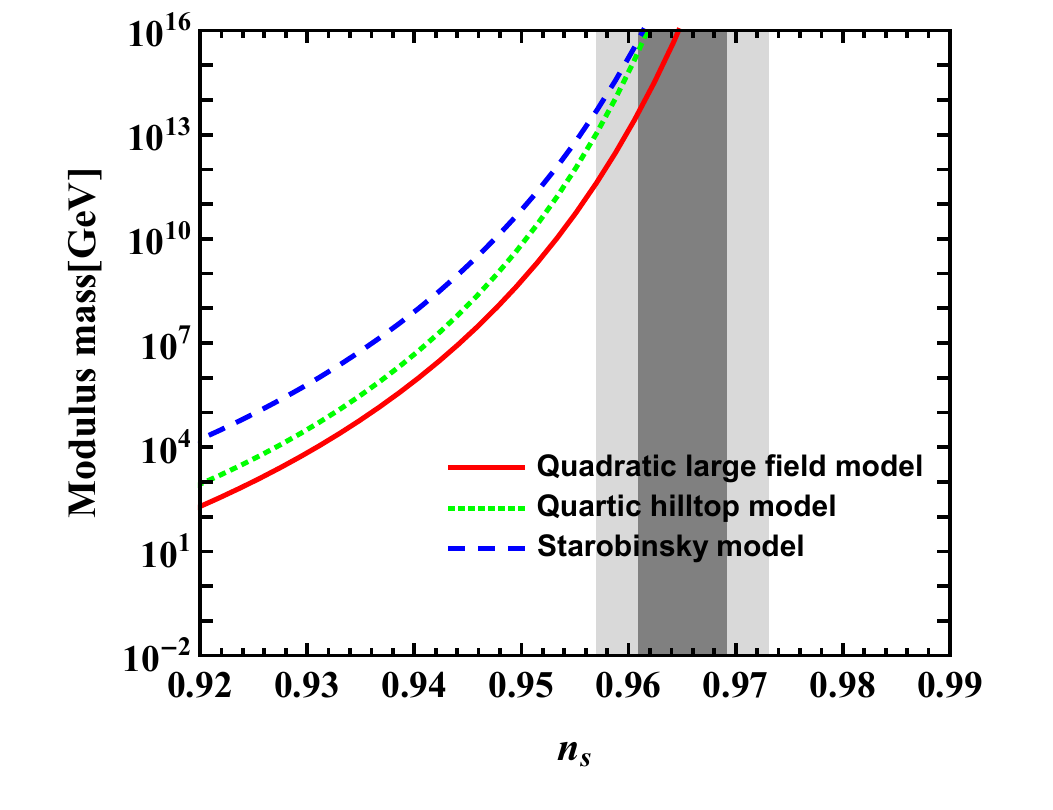}
\caption[Constraints on modulus mass]{Plots of allowed modulus mass $m_{\chi}$ 
as a function of $n_{s}$ for different inflationary models by considering the 
constraints of Ref. \cite{Goswami:2018vtp} obtained for a successful 
explanation of the CMB low multipole anomalies by considering a step in the 
inflaton potential. The solid red, dotted green and dashed blue curves 
represent the quadratic large field, quartic hilltop, and Starobinsky model 
respectively. The dark gray and light gray shaded regions correspond to the 
1$\sigma $ and 2$\sigma$ bounds respectively on $n_{s}$ from Planck 2018 data 
(TT, TE, EE + lowE + lensing) 
\cite{Aghanim:2018eyx}.}
 \label{ch9fig4}
\end{figure}

In Ref. \cite{Goswami:2018vtp} we showed that the successful explanation of the 
CMB low multipole anomalies by considering a feature in the inflaton potential 
the reheating parameters $T_{\text{reh}}$, $\bar{w}_{\text{reh}}$ and 
$N_{\text{reh}}$ can be constrained for the large field, quartic hilltop and 
Starobinsky models. Using these constraints on the reheating parameters (the 
upper bounds on $\bar{w}_{\text{reh}}, T_{\text{reh}}$), we plot  the modulus 
mass as a function of the scalar spectral index $n_{s}$ for different single 
field inflationary models in Fig. \ref{ch9fig4}. For Planck's $2\sigma$ lower limits of
$n_{s}$, we obtain $m_{\chi}\approx 4.06\times 10^{11}, 1.08\times 
10^{13}$ and $4.83\times 10^{13}$ GeV respectively for the quadratic large 
field, quartic hilltop and Starobinsky model. The $1\sigma$ lower limit of 
$n_{s}$ gives $m_{\chi}\approx 4.74\times 10^{13}$GeV, $2.84 \times 10^{15}$GeV and 
$5.94\times 10^{15}$ GeV for the quadratic large field, quartic hilltop and 
Starobinsky model respectively. The $1\sigma$ and $2\sigma$ upper limits of $n_{s}$ give 
$m_{\chi}> M_{\text{Pl}}$ which rules 
out late time modulus cosmology.

\subsection{Effects of including $N_{\text{eff}}$ in the analysis}
\label{sec:includeNeff}
As reported by Planck 2018 \cite{Akrami:2018odb}, the results for the tensor to scalar ratio $r$ are model dependent.
The alternative ranges for $n_s$ in conjunction with variation in other parameters are reported in their
Table 3. The case that substantially deviates from the others and impacts our results the most corresponds to the inclusion of 
the effective  number of neutrinos, $N_{\text{eff}}$ in the analysis.  
For instance including $N_{\text{eff}}$ in the case of Planck TT,TE,EE +lowEB+lensing allows $N_{\text{eff}}=2.92\pm 0.19$ while  
$n_s=0.9607$ with error range $( -0.0084, +0.0086)$. 
The $1\sigma$ lower limit of this  $n_{s}$ gives $m_{\chi}\approx 4.20\times 10^{9}$GeV, $5.78 \times 10^{10}$GeV and 
$4.72\times 10^{11}$ GeV for the quadratic large field, quartic hilltop and 
Starobinsky model respectively. Hence this reduces all our answers for $m_{\chi}$ by almost $4$ orders of magnitude. 
The central value of this $n_{s}$ predicts the value of $m_{\chi}$ $\approx$ $3.21\times 10^{13}$GeV, $1.81 \times 10^{15}$GeV and 
$4.0\times 10^{15}$ GeV for the quadratic large field, quartic hilltop and  Starobinsky models respectively.

\section{Discussion and conclusions}\label{Sec6}
We have obtained a relation among the reheating parameters, the 
modulus mass $(m_{\chi})$ and the inflationary observables for single field inflationary models.
This is done by tracing the evolution of specific observable scales in CMB 
from the time of their Hubble crossing during inflation to the present  time. 
This is in principle the same methodology as followed in \cite{Dutta:2014tya} which
obtained a bound $m_{\chi}>10^9$ GeV by obtaining a relation between $m_\chi$
and scalar spectral index $n_s$. However, by carefully pursuing the analysis
and insisting that $\bar{w}_\text{reh}$ lies within physically acceptable
range and $T_{\text{reh}}$ within the phenomenologically constrained range, we
obtain a substantially stronger bound.
This is shown in Fig.s \ref{ch9fig1}, \ref{ch9fig2}, and \ref{ch9fig3}.
Generically our lower bounds on $m_{\chi}$ for all models are high, $\gtrsim10^{12}-10^{15}$GeV.
Conversely, for $\bar{w}_{\text{reh}}<\frac{1}{3}$, the lowest value of reheat temperature we get 
is $T_{\text{reh}}\gtrsim10^5$ GeV obtained by requiring that the modulus mass remains 
sub-Planckian, $m_{\chi}<M_{\text{Pl}}$.
Further, we considered modeling the CMB low multipole anomalies 
through feature in the inflaton potential. This gives a further handle on the reheat 
parameters. The simultaneous demand of explaining the CMB 
anomalies and disappearance of the heavy moduli to accord with observed 
cosmology requires that $10^{13}~\text{GeV} \lesssim m_{\chi}\lesssim 10^{15}$ 
GeV depending on the model. 
An exception to these results arises if, as discussed in Sec. \ref{sec:includeNeff}, one include $N_{\text{eff}}$ 
parameter along with the $\Lambda$CDM model in  the analysis to determine $n_{s}$.  This reduces 
all our bounds on $m_{\chi}$ by almost $4$ orders of magnitude.

Thus we have leveraged the requirement that the Universe remain Friedmann like, 
homogeneous and  isotropic at all epochs including the reheating phase. This allows us to parameterise
the transitory epoch by an effective equation of state parameter for the material content,
agnostic of the particle physics details of how this reheating proceeds. The fact that the physically
permitted range for this parameter is highly restricted, leads to rather stringent constraints
on new physics that may intervene, such as the moduli fields. While a fully detailed model of reheating may 
provide even more refined information we do have reasonably robust relation of the light moduli 
mass on other physical parameters. The study may be taken as a demonstration that substantial 
knowledge about an epoch such as the reheating phase buried deep in the early epochs of the Universe 
is accessible by CMB observables today. 

\section{Appendix : Reheating parameters: extension to modulus dominated  case}
\label{sec2}
We begin with recapitulating the essentials of the formalism. 
The inflaton $\phi$ is considering to be governed by a potential $V(\phi)$ 
undergoing 
slow-roll evolution 
with parameters $\epsilon$ and $\eta$, resulting in scalar curvature power 
spectrum $P_\zeta(k)$ and
tensor power spectrum $P_h(k)$ as a function of the Fourier transform variable 
$k$ of the argument 
of the spatial correlation functions, with corresponding indices $n_s-1$ and 
$n_T$. The details of 
the definitions and notation are standard \cite{Liddle:1993fq}, and can be 
found
also in 
the references \cite{Riotto:2002yw,Bassett:2005xm,Martin:2013tda}.
We shall use $A_{s}$ and $A_{T}$, the amplitude of scalar and tensor power 
spectra at 
the pivot scale $k_{*}$ as used by Planck collaboration, 
$\frac{k_{*}}{a_{0}}=0.05 
\text{Mpc}^{-1}$. For  $k=k_{*}$, these amplitudes are given in terms of 
$H_{*}$ 
as 
\be\label{e7}
A_{T}= P_{h}(k_{*})=\frac{2H_{*}^{2}}{\pi^{2}M_{\text{Pl}}^{2}}, \hspace{1cm} 
A_{s}= P_{\zeta}(k_{*})=\frac{H_{*}^{2}}{8\pi^{2}M_{\text{Pl}}^{2}\epsilon_{*}}.
\ee
In terms of the slow-roll parameters $\epsilon$ and $\eta$, the tensor to 
scalar 
ratio $r$, the 
scalar spectral index $n_{s}$ and the tensor spectral index 
$n_{T}$ satisfy the relations 
\be\label{ns}
 r=16\epsilon,\hspace{1cm} n_{s}=1-6\epsilon+2\eta,\hspace{1cm} 
n_{T}=-2\epsilon.
\ee
The total number of e-foldings, $N_{T}$, is defined as the logarithm of the 
ratio of the scale factor at the final time $t_{e}$ to it's value at initial 
time $t_{i}$ of the era of inflation.
\be
 N_{T}\equiv\ln\frac{a(t_{e})}{a(t_{i})}=\int_{t_{i}}^{t_{e}}{H 
dt}=\int_{\phi_{i}}^{\phi_{\text{end}}}{\frac{H}{\dot{\phi}}d\phi}=\frac{1}{M_{
\text{Pl}}}\int^{\phi_{i}}_{
\phi_{\text{end}}}{\frac{1}{\sqrt{2\epsilon}}d\phi}.
\ee
Where $\phi_{i}$ and $\phi_{\text{end}}$ are the initial and final values of 
the 
inflaton field $\phi$ and 
$\epsilon$ is the slow-roll parameter defined as 
$\epsilon=-\dot{H}/H^{2}=\dot{\phi}^2/(2H^2 
M_\mathrm{Pl}^2 )$.
Likewise, given a mode $k$, the number of e-foldings between the time when 
it crosses the Hubble 
horizon and the end of inflation is given by
\be\label{NKE}
 \Delta 
N_{k}=\int_{\phi_{k}}^{\phi_{\text{end}}}{\frac{H}{\dot{\phi}}d\phi}=\frac{1}{M_
{\text{Pl}}}\int^{\phi_{k}}_{
\phi_{\text{end}}}{\frac{1}{\sqrt{2\epsilon}}d\phi},
\ee
where $\phi_{k}$ is the value of the inflaton field at the time of Hubble 
crossing of the scale k. For the 
slow-roll approximation i.e., $V(\phi)\gg\dot{\phi}^{2}$ and 
$\ddot{\phi}\ll3H\dot{\phi}$ the Eq. \eqref{NKE} 
becomes 
\be\label{e11}
 \Delta 
N_{k}\approx\frac{1}{M_{\text{Pl}}^{2}}\int_{\phi_{\text{end}}}^{\phi_{k}}{\frac
{V}{V'}d\phi}.
\ee
\par
In the inflationary model of cosmology, at the end of the inflation, the 
inflaton field decays and reheat the Universe. The energy stored in the inflaton 
gets converted to radiation. The Hubble parameter $(H)$ and the energy density 
of the Universe decrease with the expansion of the Universe. When the Hubble 
parameter’s value becomes equivalent to the mass of a modulus, the modulus 
field start oscillating in its potential minimum
\cite{Dutta:2014tya,Das:2015uwa,Coughlan:1983ci,Banks:1993en,deCarlos:1993wie}.
The energy density of the modulus field redshifts like matter (which is slower 
than the redshift rate of radiation); hence, the energy density of 
the Universe becomes modulus dominated. After this, the modulus decays and 
reheat the Universe for the second time. During inflation, The equation of 
motion of a scalar field $\chi$ is given by
\be
\ddot{\chi}+(3H+\Gamma_\chi)\dot{\chi}+\frac{\partial V}{\partial\chi}=0.
\ee
Where $\Gamma_{\chi}$ is the decay width of the scalar field $\chi$ and $H$ is 
the Hubble parameter. If the value of the Hubble parameter is greater than 
the mass of the scalar, $m_{\chi}$, then the field will freeze at its initial 
displacement $\chi_{\text{in}}$. This initial displacement is of the order of 
$M_{\text{Pl}}$. The quantum fluctuation \cite{Goncharov:1984qm} of the field 
during inflation or the dependence of the modulus potential on the vacuum 
expectation value \cite{Dine:1983ys, Coughlan:1984yk, 
Dine:1995uk, Dine:1995kz, Dutta:2014tya} of the inflaton are the possible 
reasons for this 
initial displacement of the scalar field. The modulus starts oscillating around 
its minimum, and then the energy density of modulus (matter) and radiation 
becomes equal, and is given by  
\cite{Dutta:2014tya}
\be\label{ch9rhoeq}
\rho_{\text{eq}}=m_{\chi}^{2}\chi_{\text{in}}^{2}\left(\frac{\chi_{\text{in}}^{2
} } { 6M_ { \text {Pl}}^{2} } \right)^3.
\ee
The energy density of the modulus then dominates, and the modulus decays at 
energy density
\be\label{ch9decay}
\rho_{\text{decay}}\sim M_{\text{Pl}}^{2}\Gamma_{\chi}^{2}.
\ee
The lifetime of the modulus $\tau_{\text{mod}}$ is expressed as
\be\label{ch9tau}
\tau_{\text{mod}}\approx\frac{1}{\Gamma_{\chi}}\approx\frac{16\pi 
M_{\text{Pl}}^{2}}{m_{ \chi}^{3}}
\ee
Using Eqs. \eqref{ch9decay} and \eqref{ch9tau} the  temperature after the modulus decay can be 
written in terms of the modulus mass as 
\be\label{ch9reh}
T_{\text{decay}}\sim m_{\chi}^{3/2}M_{\text{Pl}}^{-1/2}
\ee
The lower bound of the reheat temperature is around a few MeV (the BBN 
temperature). Hence, using Eq. \eqref{ch9reh} one obtains the bound on the 
modulus mass (known as cosmological moduli problem bound) as 
\cite{Coughlan:1983ci,Banks:1993en,deCarlos:1993wie}
\be
m_{\chi}\ge30 \text{TeV}
\ee
\par

\bibliographystyle{apsrev}
\bibliography{ref}
\end{document}